\documentclass[12pt, draftclsnofoot, onecolumn]{IEEEtran}
\usepackage{balance}
\usepackage{cite, graphicx,amssymb,color,  pict2e, multirow, rotating} 
\usepackage{graphicx,amssymb}
\usepackage[tbtags]{amsmath}
\usepackage{times,amsmath}
\usepackage{subfig}
\usepackage{amsmath}
\usepackage{epsfig}
\usepackage{amssymb}
\usepackage{hyperref}
\usepackage{color}
\usepackage{psfrag}
\usepackage{subfig}
\usepackage[usenames,dvipsnames]{xcolor}
\usepackage{textcomp}
\usepackage{array}
\usepackage{tikz}
\usetikzlibrary{matrix,decorations.pathreplacing}
\usepackage{booktabs}

\usepackage{xr}

\usepackage{tabularx,ragged2e}
\newcolumntype{C}{>{\Centering\arraybackslash}X} 

\usepackage{algorithmic}
\usepackage{algorithm}

\newtheorem{definition}{Definition}

\newtheorem{lemma}{Lemma}

\newtheorem{theorem}{Theorem}

\begin{document}
%
\title{MAC for Machine Type Communications in Industrial IoT -- Part II: Scheduling and Numerical Results}
%
%
%

\author{
	    Jie Gao,~\IEEEmembership{Member,~IEEE,}
	    Mushu Li,~\IEEEmembership{Student Member,~IEEE,}
        Weihua Zhuang,~\IEEEmembership{Fellow,~IEEE,}
        Xuemin (Sherman) Shen,~\IEEEmembership{Fellow,~IEEE,}
        and Xu Li
\thanks{
Jie Gao is with the Department of Electrical and Computer Engineering, Marquette University, Milwaukee, WI 53233 USA (e-mail: j.gao@marquette.edu).

Mushu Li, Weihua Zhuang, and Xuemin (Sherman) Shen are with the Department of Electrical and Computer Engineering, University of Waterloo, Waterloo, ON, Canada, N2L 3G1 (e-mail:
\{mushu.li, wzhuang\}@uwaterloo.ca, xshen@bbcr.uwaterloo.ca).

Xu Li is with Huawei Technologies Canada Inc., Ottawa,
ON, Canada, K2K 3J1 (email: Xu.LiCA@huawei.com).}
}

\maketitle

\begin{abstract}
In the second part of this paper, we develop a centralized packet transmission scheduling scheme to pair with the protocol designed in Part~I and complete our medium access control (MAC) design for machine-type communications in the industrial internet of things. For the networking scenario, fine-grained scheduling that attends to each device becomes necessary, given stringent quality of service (QoS) requirements and diversified service types, but prohibitively complex for a large number of devices. To address this challenge, we propose a scheduling solution in two steps. First, we develop algorithms for device assignment based on the analytical results from Part~I, when parameters of the proposed protocol are given. Then, we train a deep neural network for assisting in the determination of the protocol parameters. The two-step approach ensures the accuracy and granularity necessary for satisfying the QoS requirements and avoids excessive complexity from handling a large number of devices. Integrating the distributed coordination in the protocol design from Part~I and the centralized scheduling from this part, the proposed MAC protocol achieves high performance, demonstrated through extensive simulations. For example, the results show that the proposed MAC can support 1000 devices under an aggregated traffic load of 3000 packets per second with a single channel and achieve $<0.5$ ms average delay and $<1\%$ average collision probability among 50 high priority devices.
\end{abstract}


\IEEEpeerreviewmaketitle

\section{Introduction}

\IEEEPARstart{I}{ndustrial} internet of things (IIoT) demands design innovations in wireless communications to enhance the support for machine-type communications (MTC)~\cite{J_YLiuProcIEEE2019}. Part~I of this work introduces our medium access control (MAC) protocol for MTC in IIoT~\cite{J_JGao_JIoT_2020PI}, which provides a potential to increase network capacity and improve quality of service (QoS) performance through increasing channel utilization efficiency. Meanwhile, how to utilize this potential to \textit{guarantee} stringent QoS requirements in a dense network calls for further investigation. Specifically, given the proposed mini-slot based slot structure and a large number of devices, proper \textit{scheduling}, i.e., determining the slot/cycle lengths and assigning the devices specific slots and mini-slots, has a significant impact on the MAC performance.

In our networking scenario, scheduling is for single-hop and uplink communications. Even in this limited scope, many research works exist in the literature, with a common focus on the trade-off between performance and signaling overhead. Early works include the development of semi-persistent scheduling for voice over IP in LTE~\cite{C_DJiang_2007}, which aims to achieve a balance between system capacity and signaling overhead. For the wireless local area network (WLAN), Wang and Zhuang propose a token-based scheduling scheme, which achieves performance prioritization for different traffic types with a low overhead in a fully connected network~\cite{J_PWangTWC_2008}. Gamage~\textit{et~al.} develop uplink scheduling for WLAN and cellular interworking to enable multi-homing voice and data services~\cite{J_AGamage_TCOM_2014}.

Despite the abundance of existing studies, scheduling in the setting of MTC and IIoT remains challenging. Ksentini~\textit{et al.} note the potentially overwhelming overhead in the uplink scheduling with a massive number of MTC connections and consider a simple round-robin scheduling algorithm for the case with no QoS requirements~\cite{M_AKsentiniNetwork_2018}. Lioumpas \textit{et al.} recognize that schedulers designed for general cellular networks cannot be directly applied to MTC, due to a higher device density and a wider variety of QoS requirements, and propose a scheduling algorithm to prioritize devices with low delay tolerance~\cite{C_ALioumpasGlobecom_2011}. However, the delay requirements considered therein is in the range from 10 milliseconds (ms) to 10 minutes, which can be too large for IIoT applications.

To handle a large number of devices, a popular strategy is to divide the devices into groups (or clusters) and schedule the devices based on the groups~\cite{J_Al-Janabi_IoT_2019}. Si~\textit{et al.} propose a grouping-based algorithm that adjusts the service rate for each user group to provide statistical QoS guarantees, where the considered delay requirements are in the range from 20ms to 100ms~\cite{J_PSi_JTVT_2014}. Karadag~\textit{et~al.} present semi-persistent scheduling for MTC in cellular networks, taking delay constraints of devices into account, where devices have periodic traffic arrivals~\cite{J_GKaradag_TWC_2019}. Zhang~\textit{et~al.} propose a random access scheme for MTC in cellular networks by grouping devices according to their delay requirements and applying access control for each group based on the group size, aggregated packet arrival rate, etc.~\cite{J_CZhangIoT_2019}. Arouk~\textit{et al.} propose a group paging based scheduling for massive MTC access in cellular networks, where the key idea is to scatter the contention for channel access to improve performance in terms of delay, collision probability, and energy consumption~\cite{J_OArouk_JSAC_2016}. The focuses of the last two works are on throughput maximization and energy consumption reduction, respectively, rather than supporting a stringent (e.g., ms level) delay requirement.

Given a high device density, diversified service types, and stringent QoS requirements, scheduling may need to be further fine-grained. Specifically, a scheduler may need to attend to the available information (e.g., packet arrival rate) or access strategy of each single device. Salodkar~\textit{et~al.} propose a learning-assisted scheduling scheme, in which each device uses reinforcement learning to determine a preferred transmission rate and a base station (BS) schedules the device with the highest rate~\cite{J_NSalodkar_2010}. Such a scheme can adapt to unknown packet arrival statistics. Chang~\textit{et~al.} propose device-level uplink scheduling schemes based on conflict-avoiding codes, in which each device is assigned a two-dimensional code matrix~\cite{J_CChang_IEEEACMNet_2019}. These schemes are applicable when multiple channels are available. In their recent work, Rodoplu~\textit{et al.} present proactive forecasting-assisted scheduling to support massive access in the internet of things (IoT), which explores machine learning to predict the traffic of each device and reserve channel time accordingly~\cite{J_VRodoplu_IoT_2020}. The scheme improves network performance with low overhead. \textcolor{black}{Yang~\textit{et~al.} utilize a neural network to predict the number of IoT devices and Wi-Fi users, which facilitates dynamic scheduling and channel allocation for co-existing IoT and Wi-Fi communications~\cite{J_BYang_TWC_2019}.}

In Part~II of this work, our objective is to develop an effective scheduling scheme to pair with the proposed protocol in Part~I. Different from the existing works, we focus on achieving QoS guarantee with very low delays. As a part of our MAC protocol, the scheduling scheme contributes to a customized link-layer solution to MTC in IIoT, supporting high device density, diversified service types, and stringent QoS targets. While we aim to maximize channel utilization efficiency through delicate \textit{distributed} coordination in the MAC protocol  in Part~I, the focus in Part~II is to develop a \textit{centralized} analysis-based scheduling scheme. The scheduling scheme should achieve a desired balance in the QoS of different services or different QoS metrics for the same service. The integration of distributed coordination and centralized control is expected to strengthen the  proposed MAC protocol.

With a large number of devices, finding a proper assignment for a centralized scheduler can be prohibitively complex. \textcolor{black}{Scheduling for a dense network with hundreds or even thousands of devices can be beyond the reach of conventional approaches, when the packet arrival rate of each device may impact the protocol parameters and the QoS requirement of each device needs to be satisfied. This motivates us to exploit neural networks to assist scheduling.} We propose to schedule in two steps, i.e., slot/mini-slot assignment and protocol parameter selection, and develop methods to reduce complexity in each step. The main contribution of this part is two-fold:
First, we develop algorithms to assign devices specific slots and mini-slots of the proposed protocol in Part~I, when the protocol parameters are given.  Based on the analytical results in Part~I, the proposed algorithms sort devices of each type, estimate the impact of potential assignments for each device, and make assignments for the devices one by one. As a result, the assignments possess the due accuracy and granularity necessary for satisfying diverse and stringent QoS requirements;
Second, to determine the protocol parameters, we exploit a deep neural network (DNN) to assist scheduling. \textcolor{black}{The DNN is structured such that it can be used given any number of devices and learn the mapping from various combinations of device and packet arrival profiles and protocol parameter settings to the resulting scheduling performance}. We demonstrate that, after sufficient training, the DNN can learn the mapping. Then, given a specific device and packet arrival profile, the DNN can be used to compare different protocol parameter settings and determine proper parameters for the proposed MAC.
In addition, we perform extensive simulations to demonstrate the properties of the proposed MAC protocol, the accuracy of the analysis in Part~I, and the performance of the scheduling scheme developed in this part.

The rest of Part~II is organized as follows. Section~\ref{s:Problem} describes the scheduling problem. Section~\ref{s:Assign} investigates the device assignment. In Section~\ref{s:DNNSchedule}, we exploit a DNN to determine protocol parameters. Section~\ref{s:Simulations} present the numerical results, and Section~\ref{s:Conclude} concludes this work.


\section{The Scheduling Problem}\label{s:Problem}

\begin{figure}
	\vspace{-1mm}
	\begin{centering}
		\includegraphics[width=0.50\textwidth]{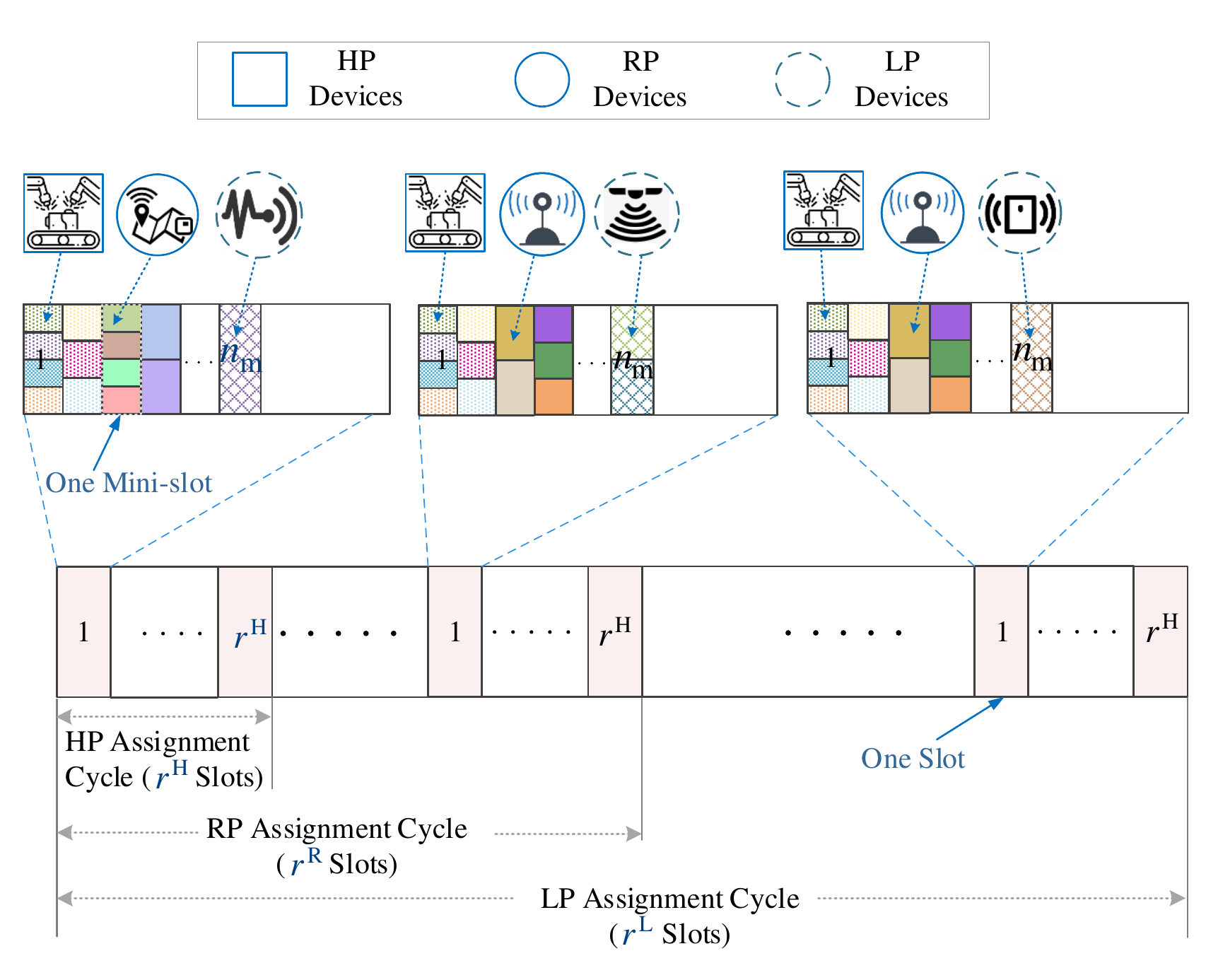}
		\par\end{centering}
	\vspace{0mm}
	\centering{\caption{An illustration of the scheduling problem. Different colors in the sub-blocks of a mini-slot correspond to different devices assigned that mini-slot, while dot-filled, solid-filled, and grid-filled patterns represent mini-slots assigned to HP, RP, and LP devices, respectively. \textcolor{black}{The scheduling problem involves determining protocol parameters $n_\mathrm{m}$, $r^\mathrm{H}$, $r^\mathrm{R}$, and $r^\mathrm{L}$ as well as assigning slots and mini-slots to all devices.} }\label{f:P2Fig1}}
\end{figure}

\textcolor{black}{Our considered network scenario and proposed MAC protocol are given in Sections~II~and~III of Part~I, respectively~\cite{J_JGao_JIoT_2020PI}. To avoid redundancy, we refer readers to the aforementioned sections for the related information.} According to the protocol description in Section~III and performance analysis in Section~IV in Part~I of this paper, it is clear that the following factors have significant impact on the performance of the proposed MAC protocol:
\begin{itemize}
	\item The number of mini-slots in each slot, i.e., $n_\mathrm{m}$;
	\item The assignment cycles $r^\mathrm{H}$, $r^\mathrm{R}$, and $r^\mathrm{L}$, which serve as different frame lengths for different types of devices;
	\item The device assignment, i.e., the allocation of devices to slots and mini-slots.
\end{itemize}
We refer to the problem of determining the above factors with the objective of satisfying QoS requirements as the packet transmission scheduling problem, which is illustrated in Fig.~\ref{f:P2Fig1}. \textcolor{black}{The access point (AP) in the network is expected to have computing capability and conduct the scheduling.} 

Note that the scheduling problem may not be always feasible. Indeed, we cannot guarantee the satisfaction of arbitrary QoS requirements given an arbitrarily large set of devices with arbitrary packet arrival rates. Thus, the objective here is to investigate effective scheduling that can support as many devices as possible while satisfying their QoS requirements.

Given the sets of all devices $\mathcal{D} = \{1, \dots, D\}$, \textcolor{black}{high-priority (HP)} devices $\mathcal{D}^\mathrm{H}$, \textcolor{black}{regular-priority (RP)} devices $\mathcal{D}^\mathrm{R}$,  \textcolor{black}{low-priority (LP)} devices $\mathcal{D}^\mathrm{L}$, and packet arrival rates $\{\lambda_i\}, i \in \mathcal{D}$, we attempt to accommodate all devices while satisfying the delay requirements $\delta^\mathrm{H}$, $\delta^\mathrm{R}$, and $\delta^\mathrm{L}$ and packet collision probability requirements $\rho^\mathrm{H}$,  $\rho^\mathrm{R}$, and $\rho^\mathrm{L}$ for the HP, RP, and LP devices, respectively. Based on the protocol, the following constraints exist for the scheduling problem (see Section~III-~D of Part~I):
\begin{itemize}
	\item The LP assignment cycle length $r^\mathrm{L}$ is a multiple of the RP assignment cycle length $r^\mathrm{R}$, which in turn is a multiple of the HP assignment cycle length $r^\mathrm{H}$;
	\item A mini-slot should not accommodate more than one type of devices;
	\item If mini-slot $m$ of slot $l$, where $l \leq r^\mathrm{H}$ and $m \leq n_\mathrm{m}$, is assigned to a subset of HP devices $\mathcal{I}^\mathrm{H}$, then mini-slot $m$ of slot $l^\prime$, for any $l^\prime \in \{r^\mathrm{H} + l, 2r^\mathrm{H} + l, \dots, r^\mathrm{L} - r^\mathrm{H} + l\}$,  is also assigned to the same set of HP devices $\mathcal{I}^\mathrm{H}$. If mini-slot $m$ of slot $l$, where $l \leq r^\mathrm{R}$ and $m \leq n_\mathrm{m}$, is assigned to a subset of RP devices $\mathcal{I}^\mathrm{R}$, then mini-slot $m$ of slot $l^\prime$, for any $l^\prime \in \{r^\mathrm{R} + l, 2r^\mathrm{R} + l, \dots, r^\mathrm{L} - r^\mathrm{R} + l\}$,  is also assigned to the same set of RP devices $\mathcal{I}^\mathrm{R}$. Both cases are illustrated in Fig.~\ref{f:P2Fig1}.
\end{itemize}

To solve the scheduling problem, we first investigate the device assignment while assuming the protocol parameters $n_\mathrm{m}$, $r^\mathrm{H}$, $r^\mathrm{R}$, and $r^\mathrm{L}$ are given. Then, we explore a DNN to assist determining these parameters. In both steps, we assume that mini-slot based carrier sensing (MsCS), synchronization carrier sensing (SyncCS), differentiated assignment cycles, and superimposed mini-slot assignment (SMsA)  from Part~I are all adopted in the proposed MAC protocol.

\section{The Device Assignment}\label{s:Assign}

In this section, we first discuss the impact of protocol parameters ($n_\mathrm{m}$, $r^\mathrm{H}$, $r^\mathrm{R}$, and $r^\mathrm{L}$) and then investigate the device assignment problem.

\subsection{Impact of $n_\mathrm{m}$}

Intuitively, increasing $n_\mathrm{m}$, subject to the conditions mentioned in the end of Section~III-B of Part~I, can support more devices via more mini-slots. However, increasing $n_\mathrm{m}$ increases delay, and consequently packet collision probability, of all devices. Therefore, given the QoS requirements of devices, increasing $n_\mathrm{m}$ may reduce the number of supported devices subject to the requirements. 

\subsection{Impact of $r^\mathrm{H}$, $r^\mathrm{R}$, and $r^\mathrm{L}$}\label{ss:CycleSlotImpact}

The delay requirements $\delta^\mathrm{H}$, $\delta^\mathrm{R}$, $\delta^\mathrm{L}$ place constraints on $r^\mathrm{H}$, $r^\mathrm{R}$, and $r^\mathrm{L}$, respectively. Consider HP devices for example. When there are $n_\mathrm{m}$ mini-slots in each slot, an upper bound on the number of slots per HP assignment cycle, i.e., $r^\mathrm{H}$, is given by~\footnote{The upper bound is obtained under the assumption that every HP device is assigned the first mini-slot of a slot. }
\begin{align}\label{e:MiniSlotTxOpRelation}
\bar{r}^\mathrm{H} = \left \lfloor{ \frac{2 \delta^\mathrm{H}}{n_\mathrm{m} T_\mathrm{m} + T_\mathrm{x}} }\right \rfloor
\end{align}
where $\left \lfloor{ \cdot }\right \rfloor$ is the floor function. \textcolor{black}{The denominator is the length of a slot. The factor `2' in the numerator follows from the fact that the average gap between the beginning of an HP cycle and the arrival of an HP packet is equal to one half of an HP cycle.} 

Using \eqref{e:MiniSlotTxOpRelation}, a relation between $n_\mathrm{m}$ and $r^\mathrm{H}$ can be obtained. If $n_\mathrm{m}$ is large,  $r^\mathrm{H}$ should be small, and the HP devices will be ``densely'' packed into the $r^\mathrm{H}$ slots. As a result, it can be challenging to satisfy the QoS requirements of HP devices. On the other hand, if $n_\mathrm{m}$ is small so that $r^\mathrm{H}$ can become large, more slots are available for HP devices in each frame. However, the transmission opportunity for RP and LP devices will decrease. Therefore, determining appropriate values for $r^\mathrm{H}$, $r^\mathrm{R}$, and $r^\mathrm{L}$ is crucial but nontrivial.

\subsection{Device Assignment}\label{ss:DeviceAssign}

The assignment of slots and mini-slots to devices is a complex problem. Consider the case with buffer and SMsA. Even if $n_\mathrm{m}$, $r^\mathrm{H}$, $r^\mathrm{R}$, and $r^\mathrm{L}$ are given, the device assignment is a combinatorial integer programming problem. Based on the analysis in Section~IV-E of Part~I, assigning any new device an occupied mini-slot can affect the delay and collision probability of all other devices assigned that mini-slot. 

We propose a heuristic algorithm for device assignment, built on the analysis in Section~IV of Part~I,  when $n_\mathrm{m}$, $r^\mathrm{H}$, $r^\mathrm{R}$, and $r^\mathrm{L}$ are given. The analysis allows us to estimate the delay and collision probability of devices in a mini-slot after adding each new device to the mini-slot. The proposed assignment algorithm tentatively assigns a device while estimating the resulting performance, with the target of satisfying the QoS requirements of all assigned devices in the process. The following settings are used in the assignment:
\begin{itemize}
	\item All devices assigned the same mini-slot have the same priority type;
	\item The maximum packet collision probability among all devices assigned the same mini-slot is referred to as the collision probability for that mini-slot and denoted by $q^\mathrm{c}_{m, l}$ for mini-slot $m$ of slot $l$;
	\item Under the assumption that the impact of collision probability on the cycle length is negligible, the length of an LP cycle can be calculated by
	\begin{align}\label{e:LPCycleLen}
	T_\mathrm{f}^\mathrm{L} = \frac{r^\mathrm{L} n_\mathrm{m} T_\mathrm{m}}{1 - \sum\limits_{i \in\mathcal{D}}\lambda_i T_\mathrm{x}},
	\end{align}
	which is based on (12) in Part~I of this paper. \textcolor{black}{The parameter $n_\mathrm{s}$ in (12) of Part~I, i.e., the number of slots in a general frame, is replaced with $r^\mathrm{L}$ in \eqref{e:LPCycleLen} since an LP cycle serves as a frame for LP devices.} Note that the use of differentiated assignment cycles does not change the packet arrival rates. \textcolor{black}{Based on the constraints mentioned in Section~\ref{s:Problem}, all devices should be scheduled at least once in an LP cycle, which leads to the summation over the packet arrival rates of all devices in the denominator of~\eqref{e:LPCycleLen}.}
\end{itemize}

Let $\hat{m}_l$ denote the minimum index among the mini-slots of slot $l$ that have not been assigned to any device. For notation simplicity, we omit subscript $l$ in $\hat{m}_l$ when $\hat{m}_l$ and $l$ both appear in the subscript (e.g., $q^\mathrm{c}_{\hat{m}_l, l}$ will be written as $q^\mathrm{c}_{\hat{m}, l}$). The length of the HP, RP, and LP assignment cycles are denoted by $T_\mathrm{f}^\mathrm{H}, T_\mathrm{f}^\mathrm{R}$, and $T_\mathrm{f}^\mathrm{L}$, respectively.  The proposed assignment is given in Algorithms~\ref{a:CoreAssign}~and~\ref{a:DeviceAssign}. Algorithm~\ref{a:CoreAssign} is the core algorithm for assigning slots and mini-slots to a set of devices with the same priority for a given cycle length, while Algorithm~\ref{a:DeviceAssign} is the overall algorithm that calls Algorithm~\ref{a:CoreAssign} to make assignments for all devices and all cycles. 

In the two algorithms, variables $n^\mathrm{c}_i$, $\Lambda_{m, l}$, and $\Gamma_{m, l}$ denote the expected number of simultaneously transmitting packets given that device $i$ is transmitting (which can be larger than 1 as a result of a nonzero collision probability), the aggregated packet arrival rate for all devices assigned mini-slot $m$ of slot $l$, and the accumulated number of packet arrivals for all devices assigned  mini-slots $1$ to $m$ of slot $l$ in the corresponding cycle, respectively. Detailed description can be found in Appendix~C of Part~I and is omitted here for brevity.

\begin{algorithm}[t!]
	\caption{Core Assignment Algorithm}\label{a:CoreAssign}
	\begin{algorithmic}[1]
		\renewcommand{\algorithmicrequire}{\textbf{Input:}}
		\renewcommand{\algorithmicensure}{\textbf{Output:}}
		\REQUIRE $\mathcal{D}^\dagger, \mathcal{R}^\dagger$, $n_\mathrm{m}$, $T_\mathrm{m}$, $T_\mathrm{x}$, $\{\lambda_i\}_{\forall i \in \mathcal{D}^\dagger}$, $r^\dagger$,  $\hat{m}_l, \forall l$, $\Gamma_{\hat{m}, l}, \forall l$.
		\ENSURE Assignment matrix $\mathbf{A}^\dagger$ with size $2 \times |\mathcal{D}^\dagger|$. \\ 
		\textit{Initialize}:  a) $q^\mathrm{c}_{m, l} =0, \forall m, l$; $n_{i}^\mathrm{c}= 0$, $\Lambda_{\hat{m},l}=0,  \forall l$;\\
		\qquad\qquad b) Number of assigned devices $N_\mathrm{a}^\dagger = 0$. \\ 
		\FOR {device $i$ in $\mathcal{D}^\dagger$}
		\STATE Check $\tau_{\hat{m}, l}, \forall l \in \mathcal{R}^\dagger$.
		\IF { $\min\limits_{l\in \mathcal{R}^\dagger}  (\tau_{\hat{m}, l} - 1) \times T_\mathrm{f}^\dagger + T_\mathrm{x} + \tau_{0}^\dagger > \delta^\dagger$}
		\STATE Quit with flag $F = i$; 
		\ELSE
		\STATE Find set $\mathcal{S}^\dagger \!=\! \{l|(\tau_{\hat{m}, l} \!-\! 1) \times T_\mathrm{f}^\dagger \!+\! T_\mathrm{x} \!+\! \tau_{0}^\dagger \leq \delta^\dagger\}$.
		\ENDIF
		\STATE Calculate $\bar{q}^\mathrm{c}_{\hat{m}, l}$ for tentative assignment $\{\hat{m}_l, l\}, \forall l\in \mathcal{S}^\dagger$, using either \eqref{e:colFirst} or \eqref{e:colUpdate} with $\tilde{q}_{m, l}^\mathrm{c}$ replaced by $\bar{q}^\mathrm{c}_{\hat{m}, l}$, depending on whether device $i$ is the first device assigned this mini-slot.
		\IF { $\min\limits_{l \in \mathcal{S}^\dagger} \bar{q}^\mathrm{c}_{\hat{m}, l}  > \rho^\dagger$ and $\hat{m}_l = n_\mathrm{m}, \forall l \in \mathcal{S}^\dagger$}
		\STATE Quit with $N_\mathrm{a}^\dagger = i$;
		\ELSIF {$\min\limits_{l \in \mathcal{S}^\dagger} \bar{q}^\mathrm{c}_{\hat{m}, l}  > \rho^\dagger$ and $\exists l \in \mathcal{S}^\dagger: \hat{m}_l < n_\mathrm{m}$}
		\STATE Update $\mathcal{R}^\dagger = \{l\in \mathcal{S}^\dagger| \hat{m}_l < n_\mathrm{m} \}$;
		\STATE Update $\hat{m}_l \!=\! \hat{m}_l \!+\! 1$, calculate $\tau_{\hat{m}, l}$, and go to Step~3;~\label{algo1:Mark1}
		\ELSE
		\STATE Find slot $l^\star =  arg\min\limits_{l \in \mathcal{S}^\dagger} \bar{q}^\mathrm{c}_{\hat{m}, l}$;
		\STATE $\mathbf{A}^\dagger(1, i) = l^\star, \mathbf{A}^\dagger(2, i) = \hat{m}_{l^\star}$;
		\STATE Update $q^\mathrm{c}_{\hat{m}_{l^\star}, l^\star}$ by setting $q^\mathrm{c}_{\hat{m}_{l^\star}, l^\star} = \bar{q}^\mathrm{c}_{\hat{m}_{l^\star}, l^\star}$;
		\STATE Update $n_{i}$, $\!\Lambda_{\hat{m}_{l^\star}\!, l^\star}$, $\!\Gamma_{\hat{m}_{l^\star}\!, l^\star}$ using \eqref{e:nFirst}~to~\eqref{e:GammaFirst} or \eqref{e:nUpdate}~to~\eqref{e:GammaUpdate}.
		\ENDIF
		\ENDFOR
		\RETURN $\{\hat{m}_l\}_{\forall l}$, $ \{\Gamma_{\hat{m}, l}\}_{\forall l}$, $\mathbf{A}^\dagger$, $N_\mathrm{a}^\dagger$.	
	\end{algorithmic}
\end{algorithm}

\begin{algorithm}[htbp]
	\caption{Overall Assignment Algorithm}\label{a:DeviceAssign}
	\begin{algorithmic}[1]
		\renewcommand{\algorithmicrequire}{\textbf{Input:}}
		\renewcommand{\algorithmicensure}{\textbf{Output:}}
		\REQUIRE $n_\mathrm{m}$, $r^\mathrm{H}$, $r^\mathrm{R}$, $r^\mathrm{L}$, $T_\mathrm{m}$, $T_\mathrm{x}$, $\mathcal{D}^\mathrm{H}$, $\mathcal{D}^\mathrm{R}$, $\mathcal{D}^\mathrm{L}$, $\{\lambda_i\}_{\forall i \in \mathcal{D}}$.
		\ENSURE Device assignment matrix $\mathbf{A}$ (size $2 \times D$),  Assignment success flag $F_\mathrm{s}$. \\  
		\textit{Initialize}: $i =1 $, $q^\mathrm{c}_{m, l} =0, \forall m, l$, $n_{j}^\mathrm{c}= 0, \forall j \in \mathcal{D}$, $F_\mathrm{s} = 0$;
		Set $\mathbf{A}^\mathrm{R}$ and $\mathbf{A}^\mathrm{L}$ to all-zero matrices with sizes $2 \times D^\mathrm{R}$ and $2 \times D^\mathrm{L}$, respectively.
		\STATE Calculate the LP Cycle length using \eqref{e:LPCycleLen}. Calculate the RP and HP Cycle length using $T_\mathrm{f}^\mathrm{R} = T_\mathrm{f}^\mathrm{L} r^\mathrm{R}/r^\mathrm{L}$ and $T_\mathrm{f}^\mathrm{H}= T_\mathrm{f}^\mathrm{L} r^\mathrm{H}/r^\mathrm{L}$, respectively.~\label{algo:MarkCycleLen}  \\
		\STATE Calculate the base delay for HP, RP, and LP devices using  $\tau_{0}^\mathrm{H} =T_\mathrm{f}^\mathrm{H}/2$,  $\tau_{0}^\mathrm{R} =  T_\mathrm{f}^\mathrm{R}/2$, $\tau_{0}^\mathrm{L} = T_\mathrm{f}^\mathrm{L}/2$, respectively. \\
		\STATE Sort devices in an increasing order of packet arrival rate for $\mathcal{D}^\mathrm{H}$, $\mathcal{D}^\mathrm{R}$, and $\mathcal{D}^\mathrm{L}$, respectively.
		\STATE Set  $\hat{m}_l = 1$, $\Gamma_{\hat{m}, l} = 0$, and $\tau_{\hat{m},l} = 1, \forall l$. Set $\mathcal{D}^\dagger = \mathcal{D}^\mathrm{H}$, $\mathcal{R}^\dagger = \{1, 2, \dots, r^\mathrm{H}\}$, $T_\mathrm{f}^\dagger = T_\mathrm{f}^\mathrm{H}$, $r^\dagger = r^\mathrm{H}$,  $\tau_{0}^\dagger=  \tau_{0}^\mathrm{H}, \delta^\dagger =  \delta^\mathrm{H}$, and $\rho^\dagger = \rho^\mathrm{H}$. Run Algorithm~\ref{a:CoreAssign} and output $\{\hat{m}_l\}_{\forall l}$, $\{\Gamma_{\hat{m}, l}\}_{\forall l}$, $\mathbf{A}^\dagger$, and $N_\mathrm{a}^\dagger$. Let $\mathbf{A}^\mathrm{H} = \mathbf{A}^\dagger$ and $N_\mathrm{a}= N_\mathrm{a}^\dagger$.~\label{algo:Mark1}\\
		\IF {$N_\mathrm{a}^\mathrm{H} = |\mathcal{D}^\mathrm{H}|$}
		\STATE Update $\hat{m}_l = \hat{m}_l + 1, \forall l$; Update $\mathcal{R}^\dagger = \{l| l \in [1, r^\mathrm{R}], \hat{m}_l \leq n_\mathrm{m}\}$; For each slot $l \in \mathcal{R}^\dagger$ and any $l^\prime \in \{r^\mathrm{H} + l, 2r^\mathrm{H} + l, \dots, r^\mathrm{R} - r^\mathrm{H} + l\}$, add $l^\prime$ to $\mathcal{R}^\dagger$ and let $\Gamma_{\hat{m},l^\prime}$ equal $\Gamma_{\hat{m},l}$. Then, calculate $\tau_{\hat{m}, l}, \forall l \in \mathcal{R}^\dagger$.~\label{algo:Mark2}  \\
		\STATE Run Algorithm~\ref{a:CoreAssign} with inputs $\{\Gamma_{\hat{m},l}\}_{\forall l}$ and $\mathcal{R}^\dagger$ from Step~\ref{algo:Mark2}, $\mathcal{D}^\dagger = \mathcal{D}^\mathrm{R}$, $T_\mathrm{f}^\dagger = T_\mathrm{f}^\mathrm{R}$, $r^\dagger = r^\mathrm{R}$,  $\tau_{0}^\dagger=  \tau_{0}^\mathrm{R}, \delta^\dagger =  \delta^\mathrm{R}$, $\rho^\dagger = \rho^\mathrm{R}$. Obtain output $\{\hat{m}_l\}_{\forall l}$, $\{\Gamma_{\hat{m}, l}\}_{\forall l}$, $\mathbf{A}^\dagger$, and $N_\mathrm{a}^\dagger$. Let $\mathbf{A}^\mathrm{R} = \mathbf{A}^\dagger$ and $N_\mathrm{a} = N_\mathrm{a} + N_\mathrm{a}^\dagger$.~\label{algo:Mark3}  \\
		\IF {$N_\mathrm{a}^\dagger = |\mathcal{D}^\mathrm{R}|$}
		\STATE Update $\hat{m}_l = \hat{m}_l + 1, \forall l$; Update $\mathcal{R}^\dagger = \{l| l \in [1, r^\mathrm{L}], \hat{m}_l \leq n_\mathrm{s}\}$; For each slot $l \in \mathcal{R}^\dagger$ and any $l^\prime \in \{r^\mathrm{R} \!+  l, 2r^\mathrm{R} \!+  l, \dots, r^\mathrm{L} \!-\! r^\mathrm{R} \!+\! l\}$, add $l^\prime$ to $\mathcal{R}^\dagger$ and let $\Gamma_{\hat{m},l^\prime}$ equal $\Gamma_{\hat{m},l}$. Then, calculate $\tau_{\hat{m}, l}, \forall l \in \mathcal{R}^\dagger$.~\label{algo:Mark4}
		\STATE Run Algorithm~\ref{a:CoreAssign} with inputs $\{\Gamma_{\hat{m},l}\}_{\forall l}$ and $\mathcal{R}^\dagger$ from Step~\ref{algo:Mark4}, $\mathcal{D}^\dagger = \mathcal{D}^\mathrm{L}$, $T_\mathrm{f}^\dagger = T_\mathrm{f}^\mathrm{L}$, $r^\dagger = r^\mathrm{L}$,  $\tau_{0}^\dagger=  \tau_{0}^\mathrm{L}, \delta^\dagger =  \delta^\mathrm{L}$, $\rho^\dagger = \rho^\mathrm{L}$. Obtain output $\mathbf{A}^\dagger$, and $N_\mathrm{a}^\dagger$. Let $\mathbf{A}^\mathrm{L} = \mathbf{A}^\dagger$ and $N_\mathrm{a} = N_\mathrm{a} + N_\mathrm{a}^\dagger$.\\
		\STATE Set  $F_\mathrm{s} = 1$ if $N_\mathrm{a} = D$.
		\ENDIF
		\ENDIF
		\RETURN $\mathbf{A} = [\mathbf{A}^\mathrm{H}, \mathbf{A}^\mathrm{R}, \mathbf{A}^\mathrm{L}]$, $F_\mathrm{s}$~\label{algo:Output}.	
	\end{algorithmic}
\end{algorithm}


\textcolor{black}{The basic ideas of Algorithms~1~and~2 are given as follows. Algorithm~1 assigns mini-slots to devices, starting from the first mini-slot of every slot, and tracks the current mini-slot being assigned. It tentatively assigns a device the current mini-slot of all available slots, trying to find the best assignment based on the resulting delay and packet collision probability estimations. If the current mini-slot in none of the slots can accommodate the device by satisfying its collision probability requirement, the algorithm moves to the next mini-slot. The procedure repeats until any of the following three conditions is satisfied: i) all devices are allocated, ii) there is no more vacant mini-slot, or iii) no current mini-slot can satisfy the delay requirement of a device. Algorithm~2 sorts the devices and calls Algorithm~1 for mini-slot and slot assignment for each device priority type. After obtaining an assignment for HP devices and RP devices, Algorithm~2 extends the assignment for the RP cycle and LP cycle, respectively.} Some details of main steps in the algorithms are summarized as follows:
 \begin{itemize}
 	\item Step~3 of Algorithm~\ref{a:CoreAssign} - The left-hand side of the inequality represents the overall delay including the base and access delays. The calculation is discussed in Section~IV-A of Part~I;
 	\item Step~\ref{algo1:Mark1} of Algorithm~\ref{a:CoreAssign} and Steps \ref{algo:Mark2} and \ref{algo:Mark4} of Algorithm~\ref{a:DeviceAssign} - These steps move from the current mini-slot to the next mini-slot of the same slot. As a result, the access delay counted in frames (AD-F) of the next mini-slot needs to be calculated. The calculation of $\tau_{\hat{m}, l}$ in these steps is based on (34) in Part~I with $T_\mathrm{f}$ replaced by the corresponding HP, RP, or LP cycle length;
 	\item Step~\ref{algo:MarkCycleLen} of Algorithm~\ref{a:DeviceAssign} - Since each LP assignment cycle consists of $r^\mathrm{L}/r^\mathrm{H}$ HP cycles and $r^\mathrm{L}/r^\mathrm{R}$ RP cycles, respectively, the HP and RP assignment cycles can be found accordingly after obtaining the LP cycle length based on \eqref{e:LPCycleLen};
	\item Step~\ref{algo:Output} of Algorithm~\ref{a:DeviceAssign} - The element in the first/second row and the $i$th column of the device assignment matrix $\mathbf{A}$ gives the index of the slot/mini-slot assigned to device~$i$;
	\item Matrix $\mathbf{A}$ only gives the first slot/mini-slot assigned to device $i$. If device $i$ is an HP device and assigned slot and mini-slot $\{l,m\}$, then it is also assigned slot/mini-slot $\{l^\prime,m\}$ for any $l^\prime \in \{r^\mathrm{H} + l, 2r^\mathrm{H} + l, \dots, r^\mathrm{L} - r^\mathrm{H} + l\}$. If device $i$ is an RP device and assigned slot and mini-slot $\{l,m\}$, then it is also assigned  slot/mini-slot $\{l^\prime,m\}$ for any $l^\prime \in \{r^\mathrm{R} + l, 2r^\mathrm{R} + l, \dots, r^\mathrm{L} - r^\mathrm{R} + l\}$. This is reflected in Steps \ref{algo:Mark2} and \ref{algo:Mark4} of Algorithm~\ref{a:DeviceAssign} and consistent with the illustration in Fig.~\ref{f:P2Fig1}.
 \end{itemize}
 In the core assignment algorithm (Algorithm~\ref{a:CoreAssign}), adding a device to a mini-slot has an impact on $\Lambda_{m,l}$,  $\Gamma_{m, l}$, and $q^\mathrm{c}_{m, l}$. Therefore, after assigning device $i$ mini-slot $m$ of slot $l$, these variables need to be updated for the mini-slot. If device $i$ is the first device assigned mini-slot $m$ of slot $l$, the following update applies:
\begin{subequations}
	\begin{align}
	\tilde{q}^\mathrm{c} _{m, l} &= 0 \label{e:colFirst} \\
	\tilde{n}_{i} &= 1,  \label{e:nFirst} \\
	\tilde{\Lambda}_{m,l} &= \lambda_{i} \\
	\tilde{\Gamma}_{m, l} &= \Gamma_{m, l} +  T_\mathrm{f}^\dagger \lambda_{i} \label{e:GammaFirst} \\
	\tilde{\tau}_{m,l} &= \tau_{m,l} \label{e:tauFirst}
	\end{align}
\end{subequations}
where $\tilde{x}$ represents an updated value of $x$ after assigning device $i$, and $T_\mathrm{f}^\dagger$ is the corresponding (HP, RP, or LP) cycle length. If device $i$ is not the first device assigned mini-slot $m$ of slot $l$, the following update applies:
\begin{subequations}
	\begin{align}
	\tilde{q}_{m, l}^\mathrm{c} &=  \left( 1 - (1 - q^\mathrm{c}_{m, l} )( 1 - T_\mathrm{f}^\dagger	\lambda_{i} ) \right)  \label{e:colUpdate} \\
	n_{i}^\mathrm{c} &= 1 +  \sum\limits_{j\in \mathcal{D}_{m, l}\backslash \{i\}} \tau_{m, l} T_\mathrm{f}^\dagger \lambda_j  \label{e:nUpdate} \\
	\tilde{\Lambda}_{m,l} &=  \Lambda_{m,l} + \lambda_{i} \bigg(1 - \frac{\tilde{q}_{m, l}^\mathrm{c}}{n_{i}^\mathrm{c}}\bigg)  \\
	\tilde{\Gamma}_{m, l} &= \Gamma_{m, l} +  T_\mathrm{f}^\dagger \lambda_{i} \bigg(1 - \frac{\tilde{q}_{m, l}^\mathrm{c}}{n_{i}^\mathrm{c}}\bigg)  \label{e:GammaUpdate}  \\
	\tilde{\tau}_{m,l} &= \tau_{m,l} \label{e:tauUpdate}
	\end{align}
\end{subequations} 
which is based on the analysis in Section~IV-E of Part~I. \textcolor{black}{The equations \eqref{e:colUpdate} to \eqref{e:GammaUpdate} update the packet collision probability~\footnote{\textcolor{black}{In practice, a guard margin may need to be applied to the estimated collision probability in \eqref{e:colUpdate}. After all, such estimation may not be sufficiently accurate since we assume no statistical knowledge of the packet arrival of any device other than the average arrival rate.}}, the average number of packets per transmission (taking collision into account), the aggregated packet arrival rate, and the accumulated number of packet arrivals, respectively, corresponding to a mini-slot after a new device is assigned that mini-slot. The last equation, i.e., \eqref{e:tauUpdate}, follows from the proof of Theorem~3 in Part~I. Specifically, the result (34) in Part~I shows that, under a low collision probability, the AD-F for devices assigned any mini-slot depends on the packet arrival rates of all devices in the preceding mini-slots, but not the packet arrival rates of other devices sharing the same mini-slot.}

%

\section{Learning-assisted Scheduling}\label{s:DNNSchedule}

The proposed device assignment in the preceding section can be applied when the parameters $n_\mathrm{m}$, $r^\mathrm{H}$, $r^\mathrm{R}$, and $r^\mathrm{L}$ are given. In this section, we propose learning-assisted scheduling to determine the values of these protocol parameters. 

\subsection{Motivation for learning-assisted scheduling}

Choosing proper values for those protocol parameters is challenging.
First, the impact of protocol parameters $n_m$, $r^\mathrm{H}$, $r^\mathrm{R}$, $r^\mathrm{L}$ and the impact of device assignment are correlated. For example, knowledge of the slot/mini-slot assignment is required to analyze the impact of $n_m$, while the assignment cannot be determined without knowing $n_m$ first.
Second, the effects of $n_m$, $r^\mathrm{H}$, $r^\mathrm{R}$, $r^\mathrm{L}$ on the performance are mutually dependent. Consider $n_m$ and $r^\mathrm{H}$ as an example. Both $n_m$ and $r^\mathrm{H}$ affect the delay of HP devices. The impact of adjusting $r^\mathrm{H}$ depends on the value of $n_m$, and the dependence is further affected by the device packet arrival rate profile. As a result, we cannot establish an analytical model for determining $n_m$, $r^\mathrm{H}$, $r^\mathrm{R}$, and $r^\mathrm{L}$. On the other hand, using brutal force  to choose their values is not viable due to the large number of diverse devices. 
There are usually too many candidate combinations of  $n_\mathrm{m}$, $r^\mathrm{H}$, $r^\mathrm{R}$, and $r^\mathrm{L}$, and each combination requires a re-calculation of the device assignment using Algorithms~\ref{a:CoreAssign}~and~\ref{a:DeviceAssign}. As the assignment algorithm is based on calculating the delay and collision probability while assigning each device, the complexity of recalculating all assignment for all combinations can be very high.~\footnote{Such complexity, as the result of a mixed integer nonlinear programming, is noted in many works, e.g., \cite{J_BYang_TMC_2020}, some of which adopt a learning-based method as a solution.}

Consequently, we  use a learning-based method to capture the impact of $n_m$, $r^\mathrm{H}$, $r^\mathrm{R}$, $r^\mathrm{L}$ and  determine their values. Specifically, we train a DNN to learn the mapping from the combination of device and packet arrival rate profiles and protocol parameter settings to the protocol performance. \textcolor{black}{A significant part of the training can be done offline to avoid a long training duration in an online setting caused by searching for and determining appropriate protocol parameters.}


%
%


\subsection{The Role of the DNN}

We use a DNN to assist in determining the parameters of the proposed MAC protocol, as follows. First, for each device and packet arrival rate profile~\footnote{We refer to the collective information including the number of HP, RP, and LP devices as well as the packet arrival rate of each device as a device and packet arrival rate profile.}, we try different combinations of $n_\mathrm{m}$,  $r_\mathrm{H}$, $r_\mathrm{R}$, and $r_\mathrm{L}$, use the heuristic algorithm to obtain the assignment, and test the resulting performance using simulations. Then, the device and packet arrival rate profile, protocol parameter settings ($n_\mathrm{m}$,  $r_\mathrm{H}$, $r_\mathrm{R}$, and $r_\mathrm{L}$), and the resulting protocol performance (as label) are used to train and test the DNN.

The data generation, training, and testing are conducted offline. When the DNN is well-trained, we can imitate the mapping from a device and packet arrival rate profile and a protocol parameter setting to the protocol performance. Accordingly, we can determine the protocol parameters online by trying different parameters on the DNN and compare the resulting performance. Recall that the packet arrival rates of devices remain constant in a relatively long duration, as mentioned in Part~I. When an update of the packet arrival rates occurs, it triggers a decision on the protocol parameters, and the DNN assists the decision making as aforementioned.    

Specifically, the DNN works as follows.
The input of the DNN includes the following two components:
\begin{itemize}
	\item Device and packet arrival rate profile - To be flexible with the number of devices, we divide the range of packet arrival rate into $I$ intervals. Letting $\lambda^{\max}$ and $\lambda^{\min}$ denote  the maximum and minimum packet arrival rates, the width of each interval is $(\lambda^{\max} - \lambda^{\min})/I$. We count the number of HP, RP, and LP devices in each of the $I$ intervals and organize the corresponding numbers into three $I \times 1$ vectors $\mathbf{c}^\mathrm{H}$, $\mathbf{c}^\mathrm{R}$, and $\mathbf{c}^\mathrm{L}$, respectively;
	\item Protocol parameter settings - The number of mini-slots in each slot ($n_m$) and the number of slots in each HP, RP, and LP assignment cycle ($r^\mathrm{H}$, $r^\mathrm{R}$, and $r^\mathrm{L}$) are the second input  component. 
\end{itemize}
The input data, $\{\mathbf{c}^\mathrm{H}, \mathbf{c}^\mathrm{R}, \mathbf{c}^\mathrm{L}, n_\mathrm{m}, r^\mathrm{H}, r^\mathrm{R}, r^\mathrm{L}\}$, is normalized by the Z-score method~\cite{J_AKumcu_JSTSP_2017} and fed to the first fully connected layer.

\begin{figure}
	\centering
	\includegraphics[width=0.42\textwidth]{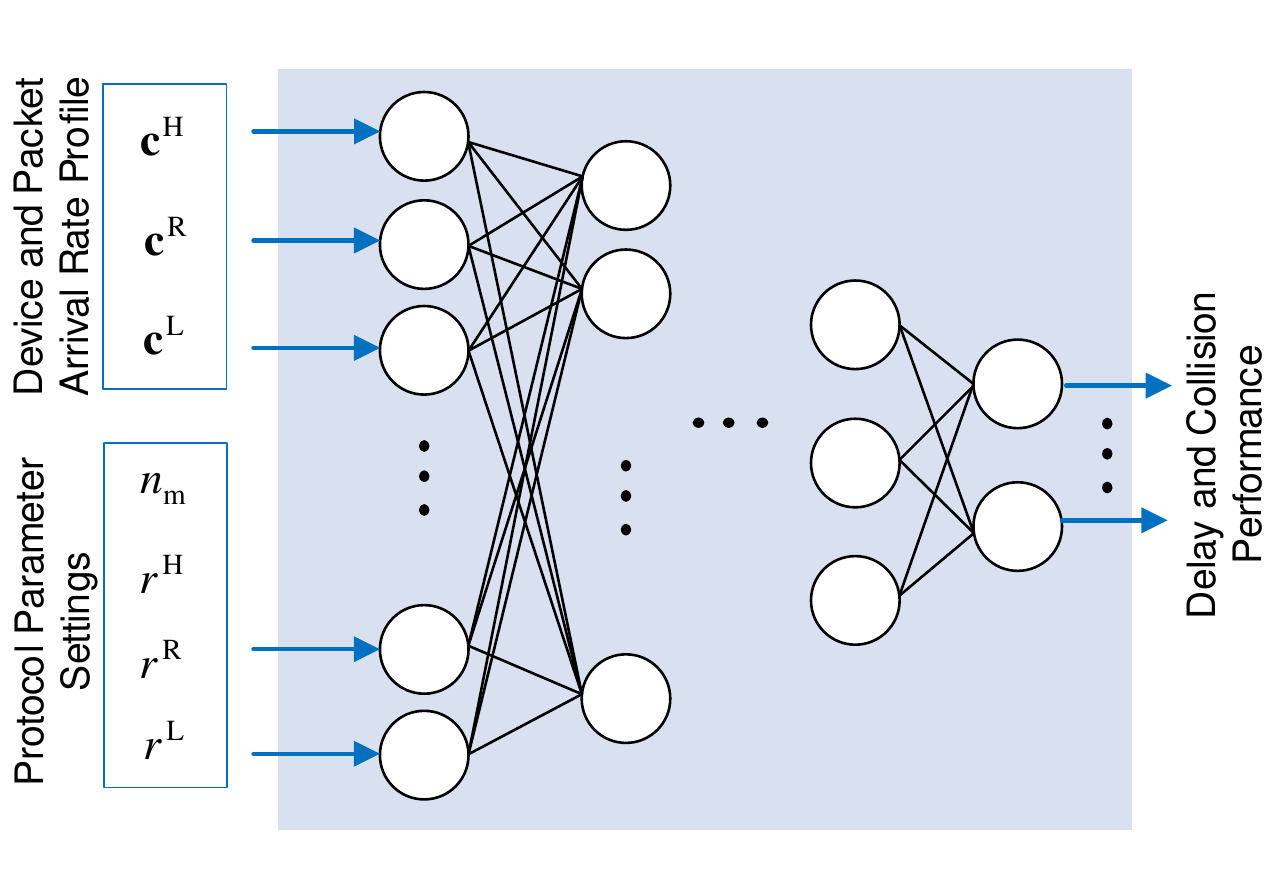}
	\caption{The Structure of the DNN.}\label{f:StructDNN}
\end{figure}

The DNN consist of $K$ fully connected layers. For layer $k$, $n_k$ neurons are deployed. The trainable parameters, i.e., kernels and bias, for neurons in the network are denoted by $\boldsymbol{\theta}$. The DNN output  includes the maximum and the average delay as well as the maximum and the average packet collision probability for each of the three device types. In addition, we adopt an indication bit in the output to indicate whether the assignment algorithms fail to find a solution that satisfies the performance requirements of all devices. The indication bit  is 1 if the assignment attempt fails and 0 otherwise. Overall, there are 13 output neurons introduced in the network.

The DNN following the above-mentioned design is illustrated in Fig.~\ref{f:StructDNN}. The DNN is implemented by Keras~\cite{O_FChollet_2019}, a high-level neural network application programming interface using Tensorflow backend. The objective of the offline training is to find an appropriate  $\boldsymbol{\theta}$ value that minimizes the loss function $\mathcal{L}(\boldsymbol{\theta})$ represented by the mean squared error (MSE)  for regression.
Adam optimizer~\cite{P_DKingma_2014} is adopted to minimize the loss function iteratively, where the optimizer is set with learning rate $\alpha = $ 1e-3 and exponential decay rates $\beta_1 = $ 0.9 and $\beta_2 = $ 0.999.

The labels, i.e., the protocol performance under specific device and packet arrival rate profiles and the protocol parameter settings, are generated via simulations. Although we can generate the labels offline, a very large training set may not be practical as it could require overwhelmingly long simulations. Meanwhile, the simulation results also demonstrate randomness, due to the randomness in the packet arrival at each device. Given the limited training set with randomness in the labels, the problem of over-fitting can be severe. We can use random dropout to alleviate over-fitting and improve the robustness of the training model~\cite{J_NSrivastava_2014}.

It is worth noting that our DNN does not directly output the best protocol parameters $\{n_\mathrm{m}, r^\mathrm{H}, r^\mathrm{R}, r^\mathrm{L}\}$. An alternative design is to train a DNN that outputs the best $\{n_\mathrm{m}, r^\mathrm{H}, r^\mathrm{R}, r^\mathrm{L}\}$. The difference is whether the DNN assists the decision making or directly makes a decision. We choose the former and let the DNN learns the mapping from various protocol parameters to the resulting performance since this approach is more flexible. For example, if the DNN directly makes a decision, the output may not be feasible or preferred when there are additional constraints on $\{n_\mathrm{m}, r^\mathrm{H}, r^\mathrm{R}, r^\mathrm{L}\}$. By contrast, using our approach, we can identify different parameter sets and compare them for a feasible or preferred solution.

\section{Numerical Results}\label{s:Simulations}

This section presents our numerical results in three parts. First, we demonstrate the effectiveness of MsCS, SyncCS, and SMsA proposed in Section~III of Part~I and verify our analysis in Section~IV of Part~I. Second, we demonstrate the performance of the device assignment in Section~\ref{s:Assign} of Part~II. Last, we demonstrate the feasibility of the DNN-assisted scheduling in Section~\ref{s:DNNSchedule} of Part~II.

The length of a mini-slot is important and should be chosen carefully. As mentioned in Part~I, the length of a mini-slot depends on the maximum propagation delay across the coverage area and the time required for detecting the channel status. The propagation time across a 500m distance, which is larger than the size of typical factories, is about $1.7 \mu$s. The channel sensing based on energy detection can be very fast and is not considered as the bottleneck for reducing the mini-slot length here~\cite{C_SYoon_Infocom2013}. However, the hardware/software incurred delay can vary for different devices. To be conservative, we use the distributed coordination function (DCF) slot time in IEEE~802.11ac as the reference and set the mini-slot time to be 9$\mu$s in most of our simulation examples~\cite{S_IEEE802.11_2016}. Using this mini-slot length, the overhead in each slot incurred by having $n_\mathrm{m}$ mini-slot for channel sensing is $9\times 10^{-6}\times n_\mathrm{m}$ seconds. For example, consider a  packet length of 50 bytes in the physical layer, and a data transmission rate of 3~Mb/s, which yields a data transmission duration of 133~$\mu$s.  With 10 mini-slots in each slot, the overall length of mini-slots is 90$\mu$s in every 233$\mu$s.


\subsection{Mini-slot Delay with MsCS, SyncCS, and SMsA}

Via simulations, we evaluate the mini-slot delay~\footnote{For brevity, we use `mini-slot delay' to refer to the delay of a device assigned that mini-slot.} in the cases with and without SyncCS and SMsA and compare the numerical results with the analytical results from Section~IV of Part~I. We focus on different mini-slots of one target slot. The general settings in this subsection are as follows (unless stated otherwise):
\begin{itemize}
	\item $n_\mathrm{m}$ and $n_\mathrm{s}$  are set to 10 and 100, respectively;
	\item $T_\mathrm{m}$ is set to $9\mu$s. $T_\mathrm{x}$ is $133\mu$s, i.e., the duration of a 50-byte physical-layer packet transmitting at 3~Mb/s. Accordingly, $T_\mathrm{s}$ in its full length is $223\mu$s, i.e., 10 $\times$ 9$\mu$s + 133~$\mu$s;
	\item Device $i$ is assigned a mini-slot with smaller index than the mini-slot of device $j$ if $\lambda_i < \lambda_j$;
	\item 20000 frames are simulated for each case.
\end{itemize}

%

\begin{figure}
	\centering
	\vspace{-2mm}
	\subfloat[0.2 to 1 packets per second per device.]
	{\includegraphics[width=0.50\textwidth]{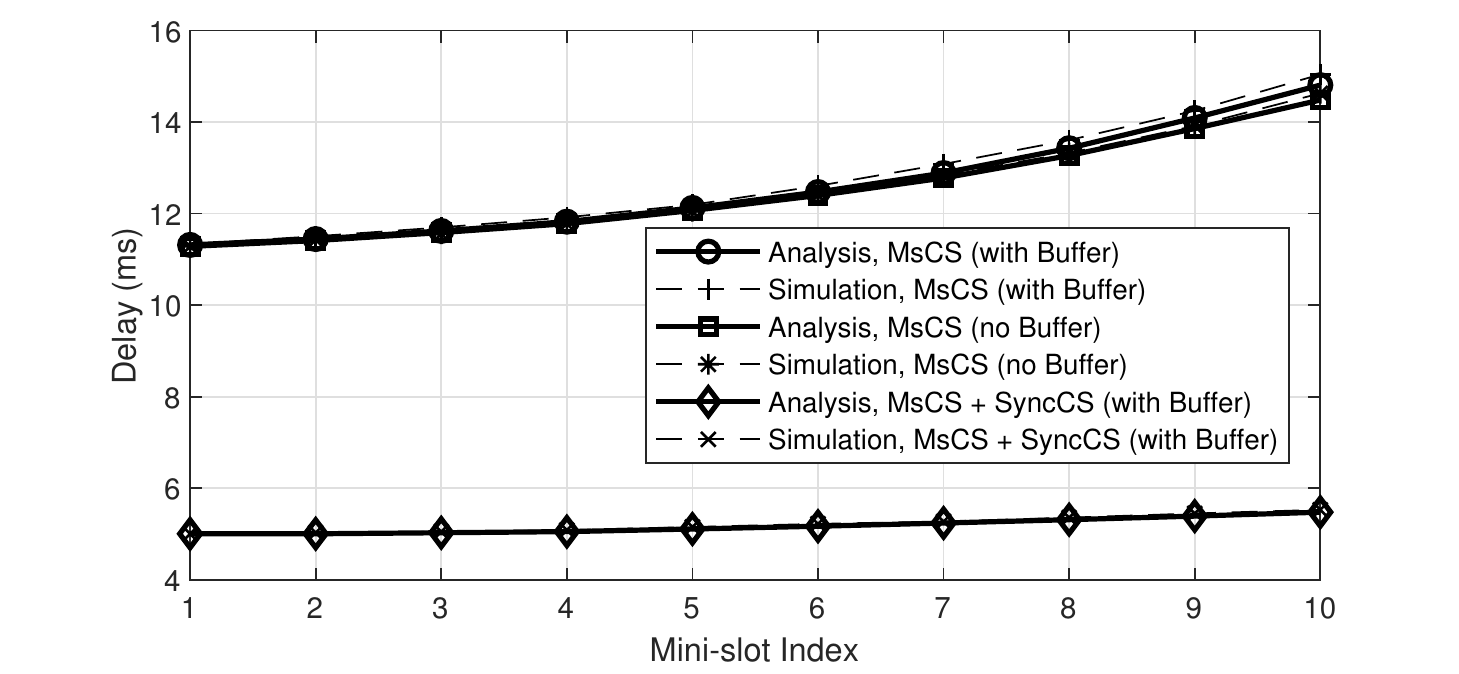}\label{f:DelayMinislot1}}
	\vspace{-1mm}
	\subfloat[1 to 5 packets per second per device.] 
	{\includegraphics[width=0.50\textwidth]{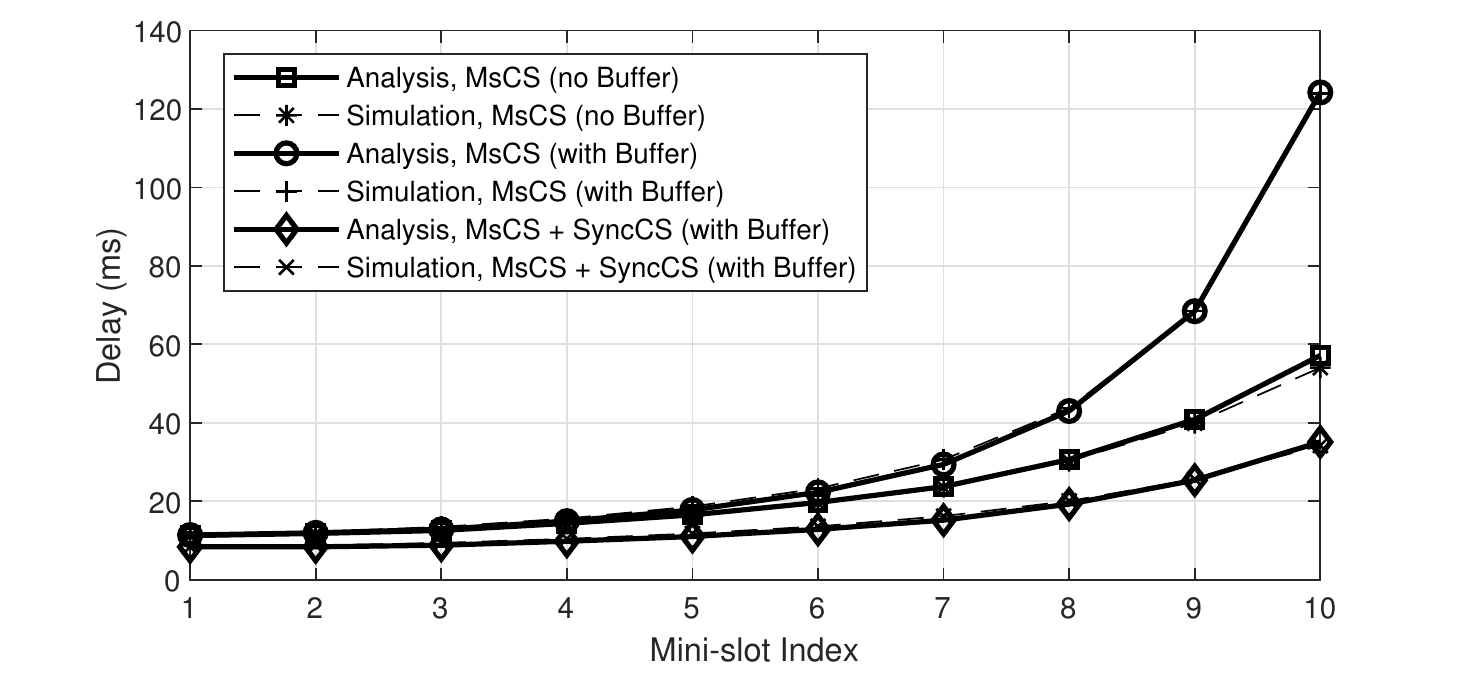}\label{f:DelayMinislot2}}
	\caption{Mini-slot delay of MsCS only and of MsCS and SyncCS with (a) lower packet arrival rates, (b) higher packet arrival rates.}\label{f:DelayMinislot}
\end{figure}

\textit{Mini-slot delay with MsCS and with MsCS and SyncCS}: Fig.~\ref{f:DelayMinislot} shows the results with only MsCS (i.e., no SyncCS or SMsA), with and without buffer, as well as the results with both MsCS and SyncCS, in the case with buffer, for Poisson packet arrivals. The overall delay includes both the base and the access delay. The packet arrival rate of each device is randomly generated based on a uniform distribution. Fig.~\ref{f:DelayMinislot}\subref{f:DelayMinislot1} corresponds to a lower packet arrival rate, i.e., in the range between 0.2 and 1 packets per second per device, and Fig.~\ref{f:DelayMinislot}\subref{f:DelayMinislot2}corresponds to a higher packet arrival rate, i.e., between 1 and 5 packets per second per device. The analytical results in Fig.~\ref{f:DelayMinislot} are based on (3) and (6) of Part~I with the expected frame length given by (14) of Part~I, respectively.  It can be observed that:
\begin{itemize}
	\item The difference between the analytical and numerical results is small for all mini-slots in all cases;
	\item The delay increases slowly with the mini-slot index for the first several mini-slots but faster for the last several mini-slots in the case of higher packet arrival rate;  
	\item The difference in delay with and without buffer is insignificant under lower packet arrival rate and significant under higher packet arrival rate;
	\item Without SyncCS, the delay for the first mini-slot is around 11ms. For the last mini-slot, depending on the packet arrival rate, the delay ranges from 15ms in Fig.~\ref{f:DelayMinislot}\subref{f:DelayMinislot1} to 125ms in Fig.~\ref{f:DelayMinislot}\subref{f:DelayMinislot2}, less than the average packet arrival interval in all cases;
	\item With SyncCS, the delay is reduced by more than $50\%$ for each mini-slot as compared with the case without SyncCS. In the case of a higher packet arrival rate in Fig.~\ref{f:DelayMinislot}\subref{f:DelayMinislot2}, the maximum delay decreases from about 125ms to around 35ms.
\end{itemize}
Overall, the numerical results demonstrate the accuracy of (3) and~(6) of Part~I,  the practicality of accommodating multiple devices in the same slot via MsCS, as well as the effectiveness of SyncCS. 

\begin{figure}
	\centering
	\vspace{-2mm}
	\subfloat[0.2 to 1 packets per second per device, no buffer.]
	{\includegraphics[width=0.50\textwidth]{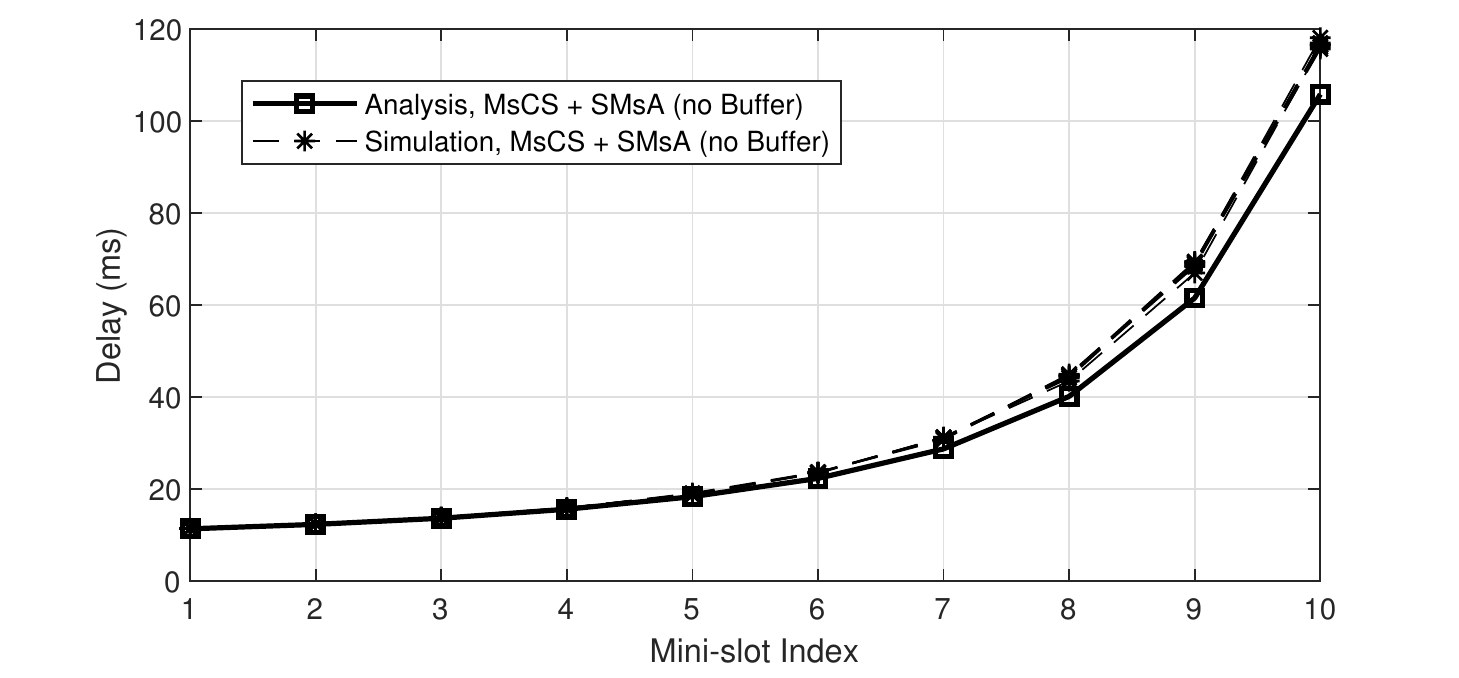}\label{f:DelayMinislotSMsA1}}
	\vspace{-1mm}
	\subfloat[0.2 to 1 packets per second per device, with buffer.] 
	{\includegraphics[width=0.50\textwidth]{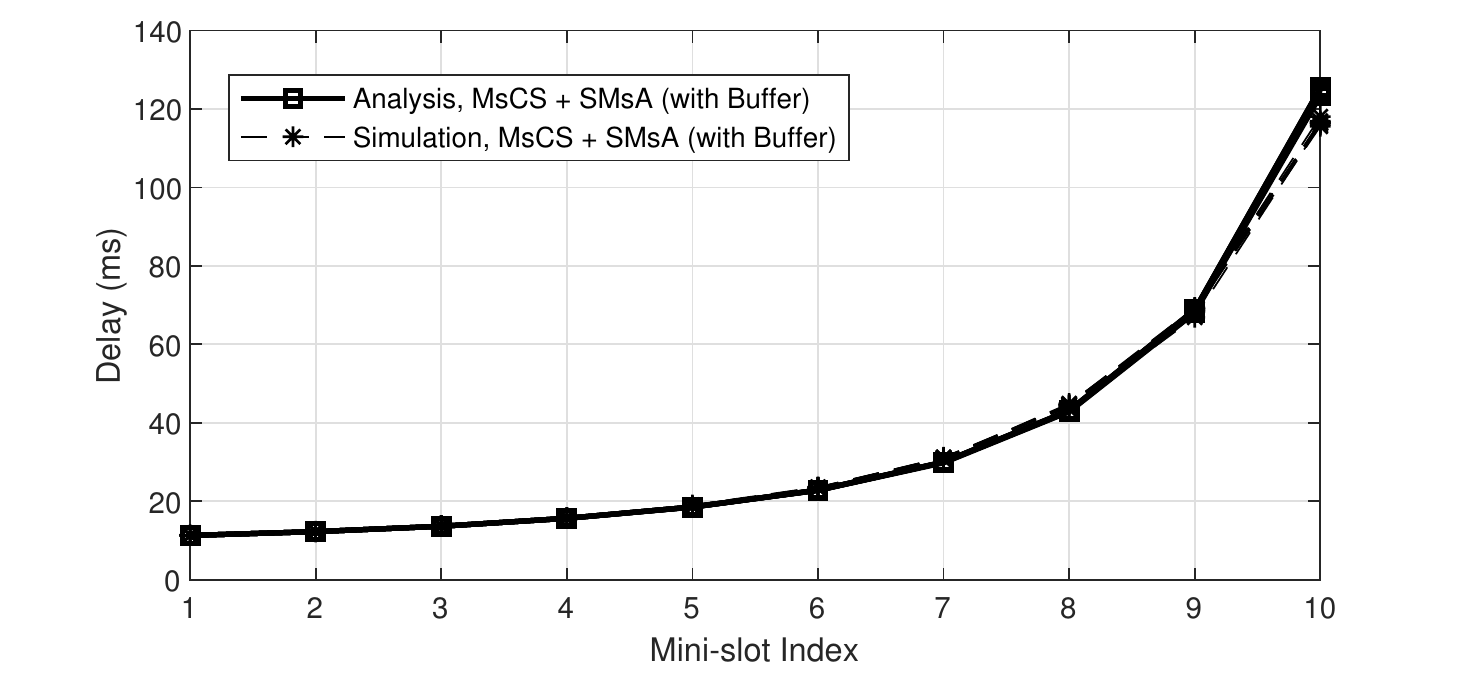}\label{f:DelayMinislotSMsA2}}
	\caption{Mini-slot delay of MsCS and SMsA with (a) no buffer, (b) with buffer. \textcolor{black}{There are 7 overlapping dashed curves in each plot, corresponding to the simulation results. Given any mini-slot index, the 7 points on the 7 dashed curves are for the 7 devices sharing the corresponding mini-slot. The only solid curve in each plot gives the analytical result for all devices, since Theorem~3 of Part~I suggests that the delay for all devices sharing the same mini-slot is approximately the same.} }\label{f:DelayMinislotSMsA}
\end{figure}

\textit{Mini-slot delay with MsCS and SMsA}: In this simulation example with SMsA (but not SyncCS), each mini-slot accommodates 7 devices instead of one. Note that such mini-slot usage is not optimal and is only used for illustrating the impact of SMsA on the mini-slot delay. As the 10 mini-slots accommodate 70 devices in total, the slot is prone to overloading if the packet arrival rate is high. Therefore, we use low packet arrival rate in this example.
Fig.~\ref{f:DelayMinislotSMsA} shows the case (a) without and (b) with buffer, respectively. Now that each mini-slot accommodates 7 devices, there are 7 numerical results on the delay for each mini-slot. The simulation results overlap in Fig.~\ref{f:DelayMinislotSMsA}, suggesting that the delay for all 7 devices in any given mini-slot is almost identical. This is consistent with Theorem~3 in Section~IV-E of Part~I. Moreover, the simulation results match closely with the analytical results based on Appendix~C of Part~I.



\begin{figure}
	\centering
	\vspace{-2mm}
	\subfloat[0.2 to 1 packets per second per device, 7$\mu$s mini-slot, 100 slots per frame.]
	{\includegraphics[width=0.50\textwidth]{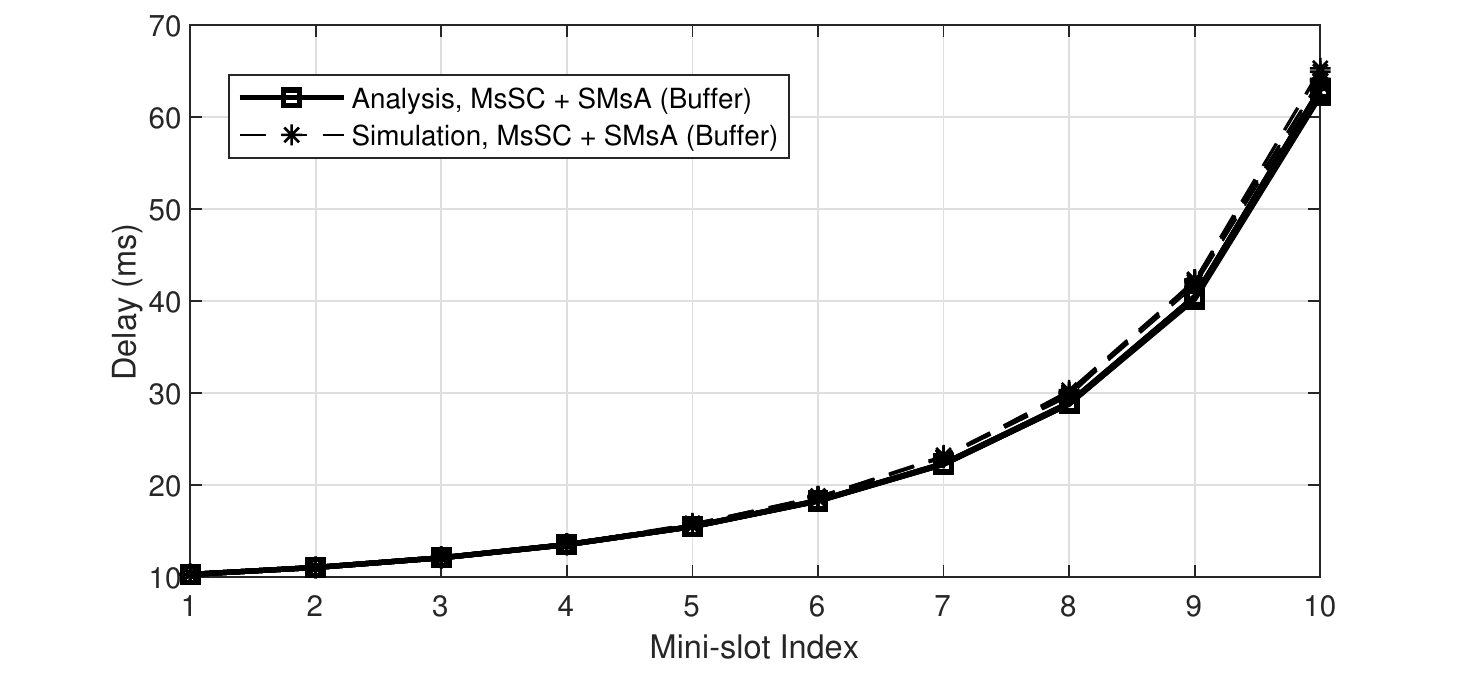}\label{f:DelayMinislotSMsASet2a}}
	\vspace{-1mm}
	\subfloat[1 to 5 packets per second per device, 9$\mu$s mini-slot, 5 slots per frame.] 
	{\includegraphics[width=0.50\textwidth]{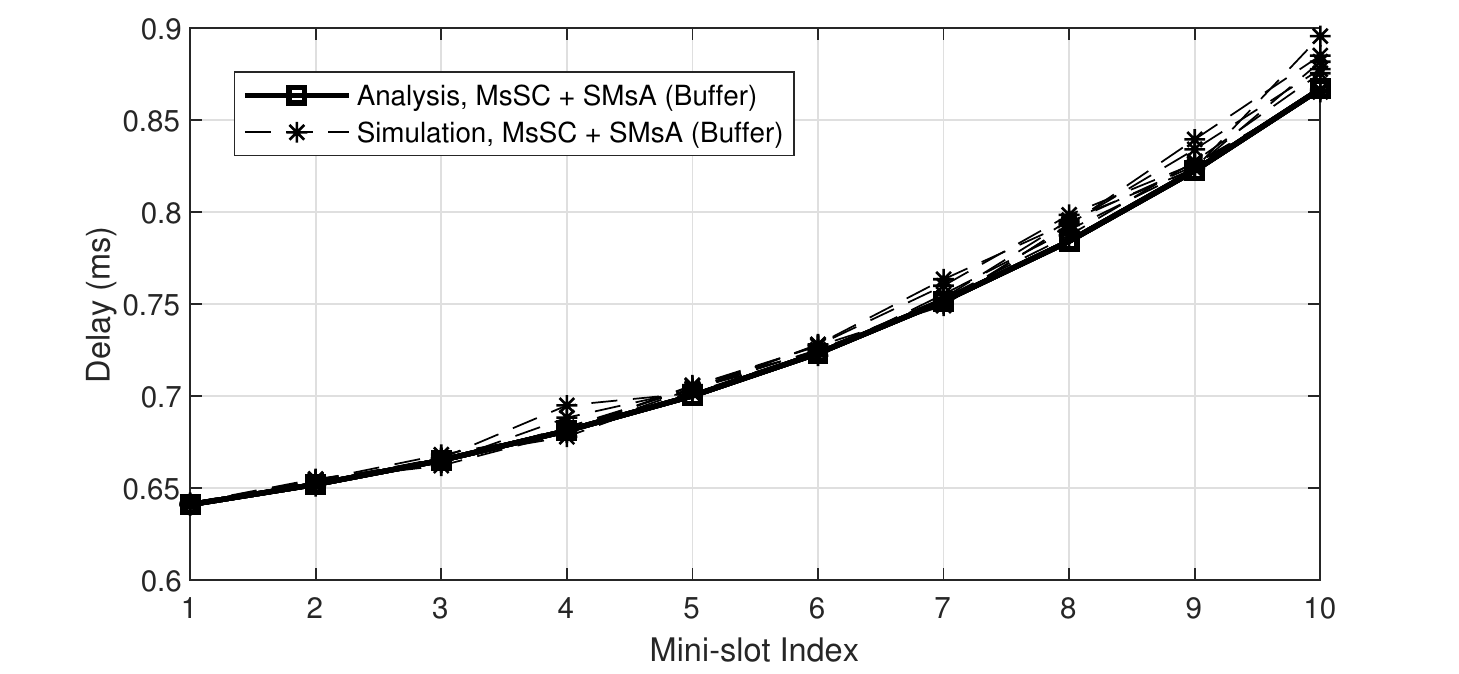}\label{f:DelayMinislotSMsASet2b}}
	\caption{Mini-slot delay of MsCS and SMsA with (a) shorter mini-slot length, (b) shorter frame length and higher packet arrival rates. \textcolor{black}{The 7 overlapping dashed curves in each plot are the result of  7 devices sharing each mini-slot. The only solid curve in each plot gives the analytical result for all devices based on Theorem~3 of Part~I.}}\label{f:DelayMinislotSMsASet2}
\end{figure}

\textit{Impact of mini-slot length and frame length}: We use the same settings as in Fig.~\ref{f:DelayMinislotSMsA} with buffers, except for a change in the mini-slot length or the frame length. The mini-slot usage here is still not optimal and only for showing the impact of mini-slot and frame lengths. In
Fig.~\ref{f:DelayMinislotSMsASet2}\subref{f:DelayMinislotSMsASet2a}, the mini-slot length reduces to $7\mu$s from  $9\mu$s in Figs.~\ref{f:DelayMinislot}~and~\ref{f:DelayMinislotSMsA}. Comparing with  Fig.~\ref{f:DelayMinislotSMsA}\subref{f:DelayMinislotSMsA2}, the impact of mini-slot length on the delay becomes evident. Accordingly, the performance of the proposed protocol can further improve if a reduction in the mini-slot length is feasible. In Fig.~\ref{f:DelayMinislotSMsASet2}\subref{f:DelayMinislotSMsASet2b}, the mini-slot length is back to $9\mu$s, the packet arrival rate is multiplied by 5, and the frame length reduces to 5 slots from 100 slots. Comparing with Fig.~\ref{f:DelayMinislotSMsA}\subref{f:DelayMinislotSMsA2}, the impact of frame length on the delay and the necessity of differentiated assignment cycles become clear. The results indicate that a very low delay is achievable if we keep the HP assignment cycle sufficiently short.

\subsection{Performance of the Device Assignment Algorithms}

We evaluate the performance of the device assignment, i.e., Algorithms~\ref{a:CoreAssign}~and~\ref{a:DeviceAssign} in Section~\ref{ss:DeviceAssign} of Part~II, given $n_m$, $r^\mathrm{H}$, $r^\mathrm{R}$, and $r^\mathrm{L}$. In the evaluation, MsCS, SyncCS, SMsA, as well as differentiated assignment cycles are used, and a buffer is assumed at each device. Again, $T_\mathrm{m}$ and $T_\mathrm{x}$ are set as $9\mu$s and $133\mu$s, respectively.

We consider 1000 devices with mixed packet arrival patterns. Specifically, the number of HP, RP, and LP devices is 50, 450, and 500, respectively. A half of all the devices, selected randomly, have Poisson packet arrivals with  rate randomly selected from the range between 1 packet per second per device and 5 packets per second per device. The remaining devices have periodic packet arrivals. The arrival rate is randomly distributed in the same range (i.e., [1, 5]), and a random component within $\pm 5\%$ of the packet arrival interval is added to each arrival instant for periodical packets. Each slot consists of 8 mini-slots (i.e., $n_\mathrm{m} = 8$), and each HP assignment cycle consists of 5 slots (i.e., $r^\mathrm{H}$ = 5). Delay thresholds $\delta^\mathrm{H}$, $\delta^\mathrm{R}$, $\delta^\mathrm{L}$ are set to 1ms, 10ms, and 80ms, respectively, while the packet collision probability thresholds $\rho^\mathrm{H}$, $\rho^\mathrm{R}$, and $\rho^\mathrm{L}$ are set to 1.5\%, 6\%, and 10\%, respectively.

\begin{figure}
	\centering
	\subfloat[Delay and collision peformance, $r^\mathrm{R}$ = 45, $r^\mathrm{L}$ = 270. ]
	{\includegraphics[width=0.5\textwidth]{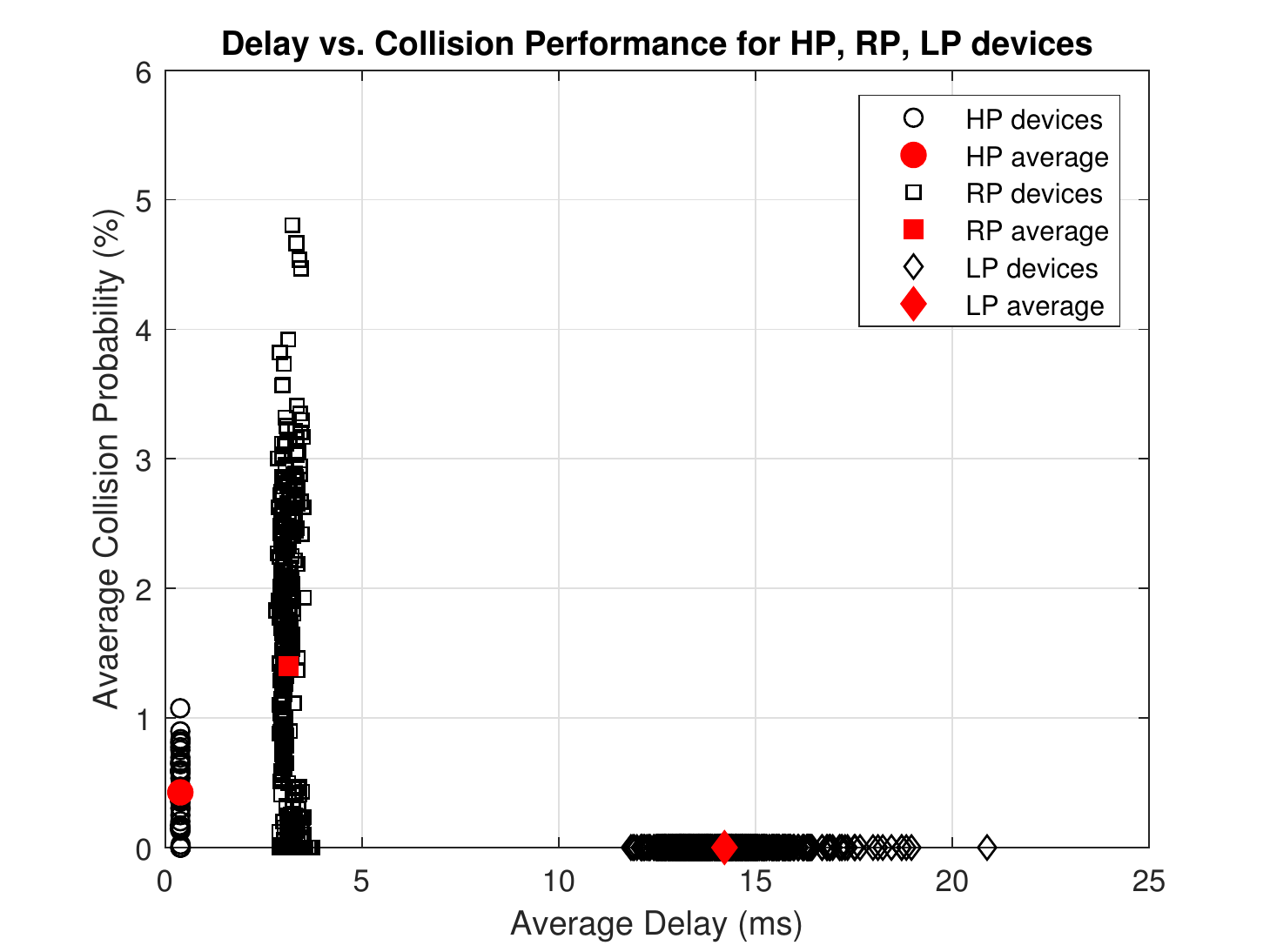}\label{f:DeviceAssign1}}
	\vspace{1mm}
	\subfloat[Delay and collision peformance, $r^\mathrm{R}$ = 35, $r^\mathrm{L}$ = 140. ] 
	{\includegraphics[width=0.5\textwidth]{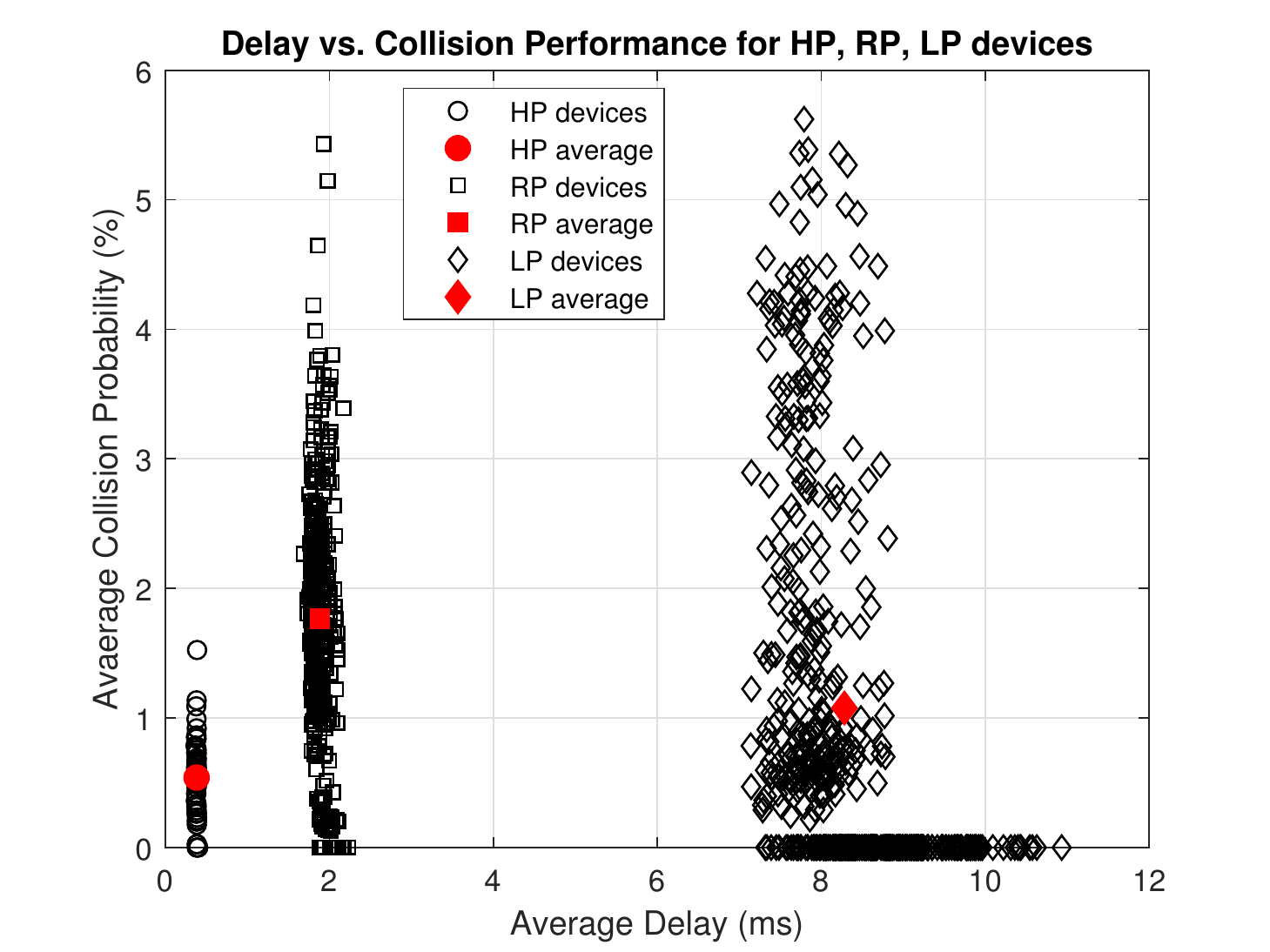}\label{f:DeviceAssign2}}
	\caption{The performance of Algorithms~\ref{a:CoreAssign}~and~\ref{a:DeviceAssign} with 1000 devices and mixed packet arrival patterns.}\label{f:DeviceAssign}
\end{figure}

A simulation duration of 2000 seconds is used to test the performance of Algorithms~\ref{a:CoreAssign}~and~\ref{a:DeviceAssign}. Fig.~\ref{f:DeviceAssign} shows the delay and packet collision probability of each device as well as the average for each type of devices, with two different assignment cycle settings. 
The three clusters in  each figure correspond to the three groups of HP, RP, and LP devices, respectively. In Fig.~\ref{f:DeviceAssign}\subref{f:DeviceAssign1}, $r^\mathrm{R}$ and $r^\mathrm{L}$ are 45 and 270, respectively, while $r^\mathrm{R}$ and $r^\mathrm{L}$ are 35 and 140 in Fig.~\ref{f:DeviceAssign}\subref{f:DeviceAssign2}. 
From Fig.~\ref{f:DeviceAssign},  we observe that the preset QoS requirements 
for all devices are satisfied. For example, from Fig.~\ref{f:DeviceAssign}\subref{f:DeviceAssign1}, 
the following observations can be made: \begin{itemize}
	\item HP devices -  average delay 0.38ms, maximum delay 0.39ms; average collision probability 0.54\%, and maximum collision probability 1.08\%;
	\item RP devices - average delay 3.1ms, maximum delay 3.7ms, average collision probability 1.4\%, and maximum collision probability 4.8\%;
	\item LP devices - average delay 14.2ms, maximum delay 20.9ms, average collision probability 0\%, and maximum collision probability 0\%.
\end{itemize}
Fig.~\ref{f:DeviceAssign} also clearly demonstrates differentiated performance achieved for different type of devices. Note that the delay in Fig.~\ref{f:DeviceAssign} is smaller than that in Figs.~\ref{f:DelayMinislot}~and~\ref{f:DelayMinislotSMsA} for two reasons. First, differentiated assignment cycles enable a very low delay for HP and RP devices. For example, each HP device gets a potential transmission opportunity in every 5 slots in the case of Fig.~\ref{f:DeviceAssign}, the same as in
Fig.~\ref{f:DelayMinislotSMsASet2}\subref{f:DelayMinislotSMsASet2b}, instead of every 100 slots in the case of Figs.~\ref{f:DelayMinislot}~and~\ref{f:DelayMinislotSMsA}. Second, each slot consists of only 8 mini-slots in the case of Fig.~\ref{f:DeviceAssign}, instead of 10 in the case of Figs.~\ref{f:DelayMinislot}~and~\ref{f:DelayMinislotSMsA}. A less number of mini-slots leads to both shorter slots, which reduce delay for all devices, and higher slot idle probabilities, which contribute to a further reduction in delay thanks to SyncCS.

Further, Fig.~\ref{f:DeviceAssign} shows the impact of assignment cycles on the performance. Specifically, via different settings of $r^\mathrm{R}$ and $r^\mathrm{L}$ in Fig.~\ref{f:DeviceAssign}\subref{f:DeviceAssign1} and Fig.~\ref{f:DeviceAssign}\subref{f:DeviceAssign2}, the possibility of making a trade-off between collision and delay is shown. Moreover, Fig.~\ref{f:DeviceAssign}\subref{f:DeviceAssign1} and Fig.~\ref{f:DeviceAssign}\subref{f:DeviceAssign2} demonstrate how our proposed algorithms can adapt to the given protocol parameters. In Fig.~\ref{f:DeviceAssign}\subref{f:DeviceAssign1}, $r^\mathrm{L}$ is larger and each LP device has to wait for a longer duration before having a transmission opportunity. As a result, the probability that an LP device has a packet to send in its assigned mini-slot can be high, and assigning two or more LP devices the same mini-slot in such case can yield a high collision probability. Therefore, the algorithms choose to assign each LP device an exclusive mini-slot. In comparison, $r^\mathrm{L}$ is much smaller in Fig.~\ref{f:DeviceAssign}\subref{f:DeviceAssign2}, and thus the probability that an LP device has a packet to send in its assigned mini-slot is lower. Therefore, the algorithms allow LP devices to share a mini-slot at the cost of small collision probabilities.

\begin{figure}
	\centering
	\includegraphics[width=0.50\textwidth]{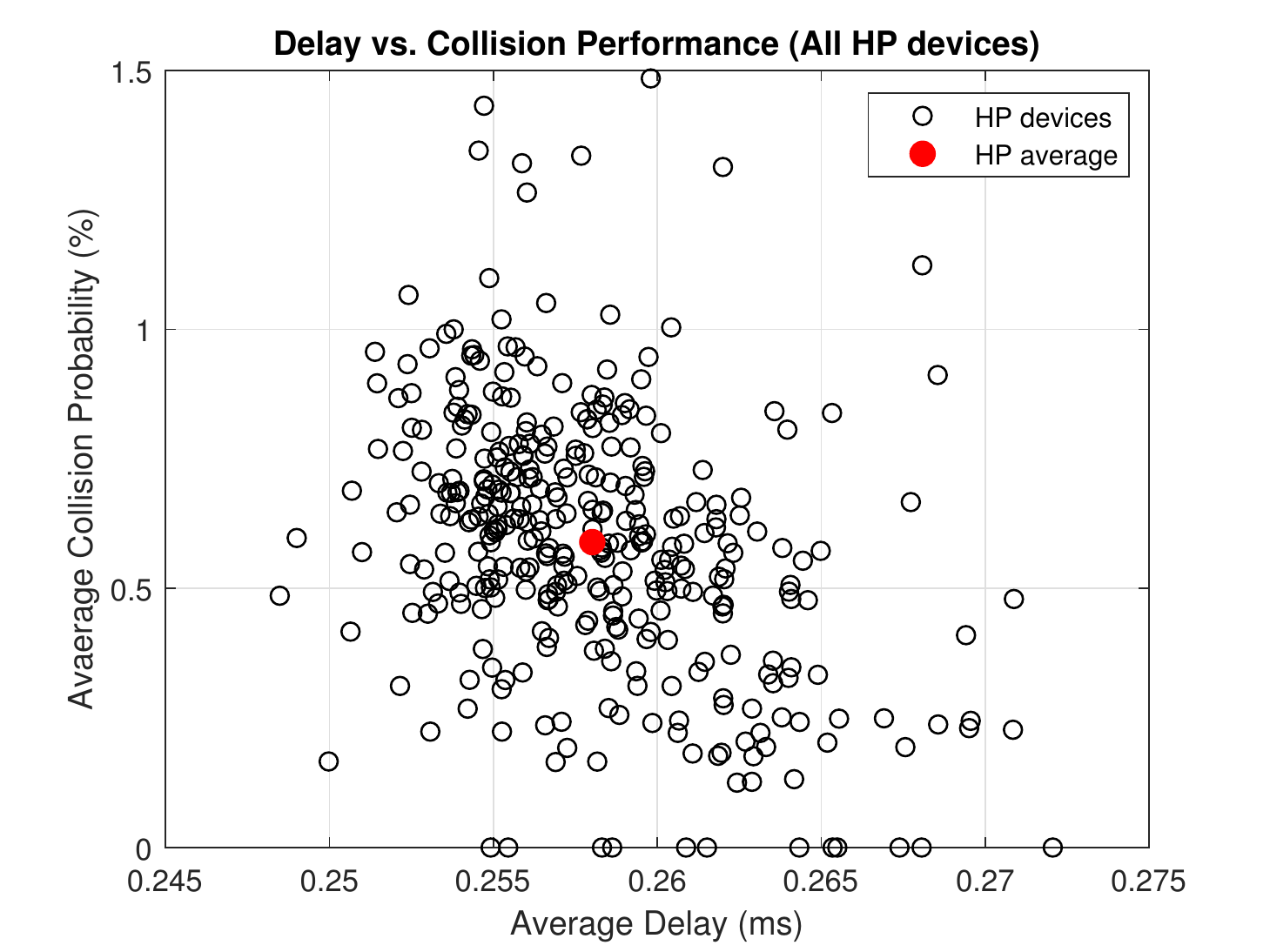}
	\caption{The performance of Algorithms~\ref{a:CoreAssign}~and~\ref{a:DeviceAssign} with 350 HP devices, $n_\mathrm{m} =4$, $r^\mathrm{H} = 6$.}\label{f:350HPDevices}
\vspace{-2mm}
\end{figure}

Fig.~\ref{f:350HPDevices} demonstrates the performance under the same setting as in Fig.~\ref{f:DeviceAssign} except: 1) there are now 350 devices, all HP, in the network; and 2) there are 4 mini-slots in each slot ($n_\mathrm{m} =4$) and 6 slots in each HP cycle ($r^\mathrm{H} = 6$). The QoS requirements on delay and packet collision are satisfied for all devices. The average delay and collision probability among all devices are less than 0.26ms and 0.6\%, respectively. This result illustrates the flexibility of the proposed device assignment algorithms in terms of adapting to different device profiles.

In the simulation examples in this subsection, the number of mini-slots, $n_\mathrm{m}$, and the assignment cycles, $r^\mathrm{H}$, $r^\mathrm{R}$, and $r^\mathrm{L}$, are not optimized. Thus, the resulting performance is not necessarily optimal. However, the results shown in Fig.~\ref{f:DeviceAssign} illustrate the advantage of the proposed MAC protocol and the assignment algorithms, in terms of satisfying stringent QoS, prioritization, and flexibility. Particularly, while random access is known to have distinctive advantage for low data traffic in delay as compared with scheduled access, e.g., as discussed in~\cite{J_MGharbieh_2018}, we demonstrate that appropriate scheduling, combined with well-designed access protocol, can also achieve very low delay in a high-density MTC network.


\begin{table}[]
	\caption{DNN Structure}\label{t.1}
	\centering{
		\begin{tabular}{@{}cccc@{}}
			\toprule
			Layer & Number of neurons & Activation function& Dropout\\ \midrule
			$n_1$ & 1024              & elu                &70\%\\
			$n_2$ & 1024              & elu                &70\%\\
			$n_3$ & 512               & elu                 &-\\
			$n_4$ & 256               & relu                &-\\
			$n_5$ & 128               & relu                &-\\
			$n_6$ & 64                & relu                &-\\
			$n_7$ & 13                & relu                &-\\ \bottomrule
	\end{tabular}}
\end{table}

\subsection{DNN-Assisted Scheduling}

The structure parameters of our proposed DNN are given in Table~\ref{t.1}. We utilize 8,200 sets of device packet arrival profiles and generate the corresponding delay and packet collision performance via the device assignment algorithms, for various values of $n_m$ and $r^\mathrm{R}$~\footnote{We fix $r^\mathrm{H}$ and $r^\mathrm{L}$ in this illustration for simplicity.}. Each of the 8200 sets consists of 6 different combinations of $n_m$ and $r^\mathrm{R}$, yielding 49,200 data entries. We employ 80\% of 49,200 data entries as the training set, 10\% as the validation set in training, and 10\% as the test set. To deal with the overfitting issue in training, we utilize the random dropout technique. Specifically, the neurons in layers $n_1$ and $n_2$ have a 70\% chance to be dropped off in each training step. The gradient backpropagation is performed over data batches of size 128  during 50 epochs.

\begin{figure}
	\centering
	\vspace{-2mm}
	\subfloat[]
	{\includegraphics[width=0.50\textwidth]{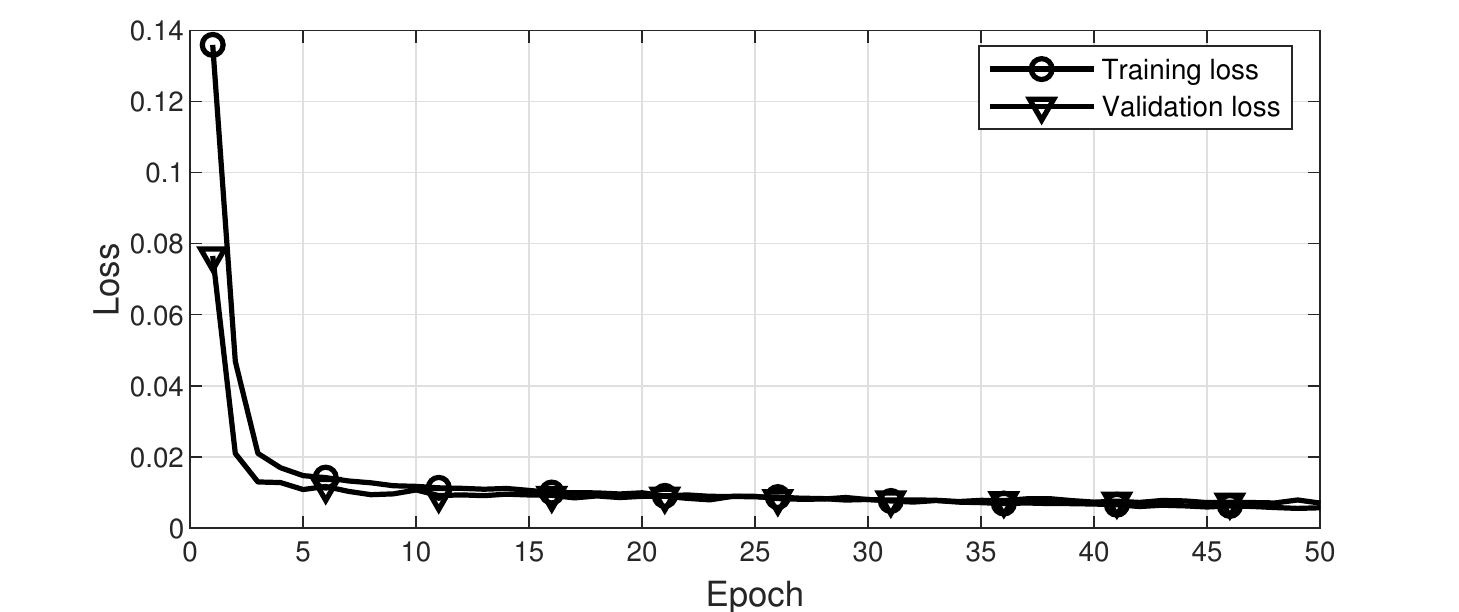}\label{f:Training1}}
	\vspace{-1mm}
	\subfloat[] 
	{\includegraphics[width=0.50\textwidth]{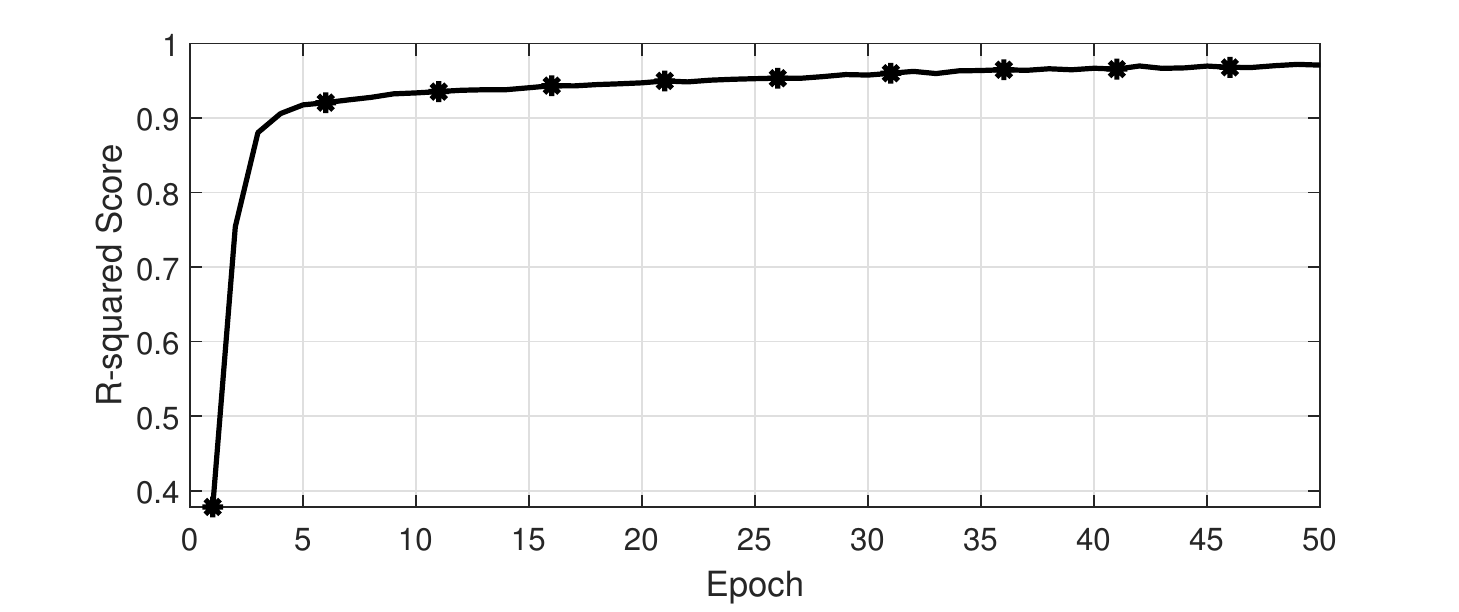}\label{f:Training2}}
	\caption{(a) Training loss and validation loss of the proposed DNN; (b) R-squared score of the proposed DNN.}\label{f:Fig11}
\end{figure}

The training loss and validation loss of the proposed DNN are shown in 
Fig. \ref{f:Fig11}\subref{f:Training1}, where the output data are normalized to the range [0, 1].  The convergence occurs after around 20 epochs. In addition, the gap between training loss and validation loss is small, showing that the overfitting issue is alleviated by random dropout.


We adopt the R-squared score to measure the fitness of our trained model in the training data set. The  R-squared score is   calculated by
\begin{equation}
\textrm{R-square} = \frac{\sum_i (\hat{y}_i - \bar{y_i})^2}{\sum_i (y_i - \bar{y_i})^2}.
\end{equation}
When the score is close to 1, the trained model can generate predicted results with a reasonably small variance. The R-squared score of the proposed DNN is shown in  Fig.~\ref{f:Fig11}\subref{f:Training2}, in which the score converges to a value close to 1 after 10 epochs.

We further validate the fitness of the trained DNN model with the data from the test set. The comparison between the predicted performance metric values and the ground truth labels is presented in Table~\ref{t.2}.~\footnote{The LP devices always have 0 collision probability in this example (similar to the case in Fig.~\ref{f:DeviceAssign}\subref{f:DeviceAssign1}). Thus, the MSE is 0 but not meaningful in such cases. Therefore, we use two `-' under LP instead of `0' in this table.}  It can be seen that the predicted results can match the ground truth labels in the test set with low MSE, and thus the proposed DNN is able to learn the mapping from the device and packet arrival profile and the protocol parameter settings to the resulting performance after sufficient training.

\begin{table}[] \caption{Comparison between predicted results and labels in the test set}\label{t.2}
	\centering
	\begin{tabular}{|c|c|c|c|c|c|c|}  \hline  \multirow{3}{*}{\begin{tabular}[c]{@{}c@{}}Overall \\ MSE\end{tabular}}      & \multicolumn{6}{c|}{Collision Probability}                        \\ \cline{2-7}                                                                          & \multicolumn{3}{c|}{Maximum  MSE} & \multicolumn{3}{c|}{Mean MSE} \\ \cline{2-7}                                                                               & HP        & LP        & RP        & HP       & LP       & RP      \\ \hline2.3e-4                                                                      & 2.9e-5   & -         & 1.8e-4   & 1.2e-6  & -        & 1.6e-5 \\ \hline\multirow{3}{*}{\begin{tabular}[c]{@{}c@{}}Flag Bit\\ Accuracy\end{tabular}} & \multicolumn{6}{c|}{Delay}                                        \\ \cline{2-7}  & \multicolumn{3}{c|}{Maximum MSE}  & \multicolumn{3}{c|}{Mean MSE} \\  \cline{2-7}  & HP        & LP        & RP        & HP       & LP       & RP      \\ \hline{98.5\%}                                                     & 5.8e-9   & 4.0e-5   & 2.6e-7   & 5.0e-9  & 1.4e-5  & 1.5e-7 \\ \hline
\end{tabular}
\end{table}

\section{Conclusion}\label{s:Conclude}

In Part~II of this paper, we customize scheduling for our proposed MAC protocol in Part~I to complete the overall MAC protocol design. To  maximize the strength of the MAC protocol, a proper choice of the cycle lengths and number of mini-slots in each slot is necessary, and so is a proper assignment of slots and mini-slots to all devices. Based on the performance analysis in Part~I, we are able to assign devices with the due granularity and accuracy. Utilizing a trained DNN, we manage to determine the protocol parameters efficiently. Integrating the distributed coordination in Part~I and the centralized scheduling in Part~II composes the unique strength of our tailored MAC design. As a result, the proposed MAC is capable of supporting a large number of devices with sporadic data packets under a single AP and a single channel, while achieving a (sub)millisecond-level delay and very low collision probability. Building on the proposed MAC, future research directions may include extending the MAC design to non-fully connected networks with either one AP or multiple APs. Another possible extension is additional transmission control measures such as random back-off or probabilistic transmission for improved fairness or further reduced collision probability.    



\section*{Acknowledgment}

The authors would like to thank Conghao Zhou and Xuehan Ye for discussions 
on the role of learning in scheduling.


%

\appendices\label{sec:Appendix}

\ifCLASSOPTIONcaptionsoff
  \newpage
\fi



\begin{thebibliography}{1}
%
%
%
%
%
%
%
%
%
%
%
%
%
%
%
%
%
%
%
%
%
%
%
%
%
%
%
%
%
%
%
%
%


\bibitem{J_YLiuProcIEEE2019}
Y. Liu, M. Kashef, K. B. Lee, L. Benmohamed, and R. Candell, ``Wireless Network Design for Emerging IIoT Applications: Reference Framework and Use Cases,''  \textit{Proc. IEEE},  vol.~107, no.~6, pp.~1166--1192, June~2019.

\bibitem{J_JGao_JIoT_2020PI}
J. Gao, W. Zhuang, M. Li, X. Shen, and X. Li, ``MAC for Machine Type Communications in Industrial IoT -- Part I: Protocol Design and Analysis,'' submitted to \textit{IEEE Internet Things J.}, under review.

\bibitem{C_DJiang_2007}
D. Jiang, H. Wang, E.  Malkamaki, and E. Tuomaala, ``Principle and Performance of Semi-persistent Scheduling for VoIP in LTE System,'' in \textit{Proc. WiCOM}, Shanghai, China, Sept.~2007, pp.~2861--2864.

\bibitem{J_PWangTWC_2008}
P. Wang and W. Zhuang, ``A Token-based Scheduling Scheme for WLANs Supporting Voice/Data Traffic and Its Performance Analysis,'' \textit{IEEE Trans. Wireless Commun.},  vol. 7, no. 5, pp. 1708--1718, May 2008.

\bibitem{J_AGamage_TCOM_2014}
A. T. Gamage, H. Liang, and X. Shen, ``Two Time-Scale Cross-Layer Scheduling for Cellular/WLAN Interworking,'' \textit{IEEE Trans. Commun.}, vol.~62, no.~8, pp.~2773--2789, Aug.~2014.

\bibitem{M_AKsentiniNetwork_2018}
A. Ksentini, P. A. Frangoudis, A. PC, and N. Nikaein, ``Providing Low Latency Guarantees for Slicing-Ready 5G Systems via Two-Level MAC Scheduling,'' \textit{IEEE Netw.}, vol. 32, no. 6, pp. 116--123, Nov./Dec. 2018.

\bibitem{C_ALioumpasGlobecom_2011}
A.~S.~Lioumpas and A.~Alexiou, ``Uplink Scheduling for Machine-to-Machine Communications in LTE-based Cellular Systems,'' in \textit{Proc. GLOBECOM Workshops}, Houston, USA,  Dec. 2011, pp. 353--357.

\bibitem{J_Al-Janabi_IoT_2019}
T. A. Al-Janabi and H. S. Al-Raweshidy, ``An Energy Efficient Hybrid MAC Protocol With Dynamic Sleep-Based Scheduling for High Density IoT Networks,'' \textit{IEEE Internet Things J.}, vol.~6, no.~2, pp. 2273--2287, Apr.~2019.

\bibitem{J_PSi_JTVT_2014}
P.~Si, J. Yang, S. Chen, and H. Xi, ``Adaptive Massive Access Management for QoS Guarantees in M2M Communications,'' \textit{IEEE Trans. Veh. Tech.}, vol.~64, no.~7, pp.~3152--3166, July 2015.

\bibitem{J_GKaradag_TWC_2019}
G. Karadag, R. Gul, Y. Sadi, and S. Coleri Ergen, ``QoS-Constrained Semi-Persistent Scheduling of Machine-Type Communications in Cellular Networks,'' \textit{IEEE Trans. Wireless Commun.}, vol. 18, no. 5, pp.~2737--2750, May 2019.

\bibitem{J_CZhangIoT_2019}
C. Zhang, X. Sun, J. Zhang, X. Wang, S. Jin, and H. Zhu, ``Throughput Optimization With Delay Guarantee for Massive Random Access of M2M Communications in Industrial IoT,'' \textit{IEEE Internet Things J.}, vol. 6, no. 6, pp. 10077--10092, Dec.~2019.

\bibitem{J_OArouk_JSAC_2016}
O. Arouk, A. Ksentini, and T. Taleb, ``Group Paging-Based Energy Saving for Massive MTC Accesses in LTE and Beyond Networks,'' \textit{IEEE J. Sel. Areas Commun.}, vol.~34, no.~5, pp.~1086--1102, May 2016.

\bibitem{J_NSalodkar_2010}
N. Salodkar, A. Karandikar, and V. S. Borkar, ``A Stable Online Algorithm for Energy-Efficient Multiuser Scheduling,'' \textit{IEEE Trans. Mobile Comput.}, vol.~9, no.~10, pp.~1391--1406, Oct.~2010.


\bibitem{J_CChang_IEEEACMNet_2019}
C. Chang, D. Lee, and C. Wang, ``Asynchronous Grant-Free Uplink Transmissions in Multichannel Wireless Networks With Heterogeneous QoS Guarantees,'' \textit{ IEEE/ACM Trans. Netw.}, vol.~27, no.~4, pp. 1584--1597, Aug.~2019.

\bibitem{J_VRodoplu_IoT_2020}
V. Rodoplu, M. Nakıp, D. T. Eliiyi and C. Güzelis, ``A Multi-Scale Algorithm for Joint Forecasting-Scheduling to Solve the Massive Access Problem of IoT,'' \textit{IEEE Internet Things J.}, to appear.
\textcolor{black}{
\bibitem{J_BYang_TWC_2019}
B. Yang, X. Cao, Z. Han, and L. Qian, ``A Machine Learning Enabled MAC Framework for Heterogeneous Internet-of-Things Networks,'' \textit{IEEE Trans. Wireless Commun.}, vol.~18, no.~7, pp.~3697--3712, July~2019.
}
\textcolor{black}{
\bibitem{J_BYang_TMC_2020}
B. Yang, X. Cao, J. Bassey, X. Li, and L. Qian, ``Computation Offloading in Multi-Access Edge Computing: A Multi-Task Learning Approach,'' \textit{IEEE Trans. Mobile Comput.}, to appear.
}
\bibitem{J_AKumcu_JSTSP_2017}
A. Kumcu, K. Bombeke, L. Platiša, L. Jovanov, J. Van Looy and W. Philips, ``Performance of Four Subjective Video Quality Assessment Protocols and Impact of Different Rating Preprocessing and Analysis Methods,'' \textit{IEEE J. Sel. Topics Signal Process.}, vol.~11, no.~1, pp.~48--63, Feb.~2017.

\bibitem{O_FChollet_2019}
F. Chollet, ``Keras: The Python Deep Learning Library,'' Astrophysics Source Code Library, 2018.

\bibitem{P_DKingma_2014} 
D. P. Kingma and J. Ba, ``Adam: A Method for Stochastic Optimization,'' arXiv:1412.6980, 2014.

\bibitem{J_NSrivastava_2014}
Nitish Srivastava, Geoffrey Hinton, Alex Krizhevsky, Ilya Sutskever, and Ruslan Salakhutdinov, ``Dropout: A Simple Way to Prevent Neural Networks from Overfitting, ''\textit{J. Mach. Learn. Res.}, vol.~15, no.~1, pp.~1929–-1958, Jan.~2014.


\bibitem{C_SYoon_Infocom2013}
S. Yoon, L. E. Li, S. C. Liew, R. R. Choudhury, I. Rhee, and K. Tan, ``QuickSense: Fast and Energy-efficient Channel Sensing for Dynamic Spectrum Access Networks,'' in \textit{Proc.IEEE INFOCOM}, Turin, Italy, Apr.~2013, pp. 2247--2255.
%

\bibitem{S_IEEE802.11_2016}
``IEEE Standard for Information Technology -- Telecommunications and Information Exchange between Systems Local and Metropolitan Area Networks—Specific Requirements - Part 11: Wireless LAN Medium Access Control (MAC) and Physical Layer (PHY) Specifications,'' in IEEE Std 802.11-2016, Dec. 2016.

\bibitem{J_MGharbieh_2018}
M. Gharbieh, H. ElSawy, H. Yang, A. Bader, and M. Alouini, ``Spatiotemporal Model for Uplink IoT Traffic: Scheduling and Random Access Paradox,''  \textit{IEEE Trans. Wireless Commun.},  vol.~17, no.~12, pp.~8357--8372, Dec. 2018.


\end{thebibliography}
%

\medskip

\balance

\bibliographystyle{IEEEtran}

%
%
%
%
%
%
%

\end{document}